\documentclass{aanew}
\usepackage{times}
\usepackage{psfig}

\def\phibar{\overline\varphi}
\def\phigeo{\varphi_{\hbox{\rm \scriptsize geo}}}
\def\phiarm{\varphi_{\hbox{\rm \scriptsize arm}}}

\def\etot{E_{\hbox{\rm \scriptsize tot}}}

\def\smp{\hskip 0.25em}

\newcommand{\psr}{PSR B0531$+$21 } 

\newcommand{\gtap}{\mathrel{\hbox{\rlap{\lower.55ex \hbox {$\sim$}}
                   \kern-.3em \raise.4ex \hbox{$>$}}}}
\newcommand{\ltap}{\mathrel{\hbox{\rlap{\lower.55ex \hbox {$\sim$}}
                   \kern-.3em \raise.4ex \hbox{$<$}}}}

\begin{document}



\title{The Crab pulsar in the 0.75-30 MeV range as seen by CGRO COMPTEL}
\subtitle{A coherent high-energy picture from soft X-rays up to high-energy $\gamma$-rays}

\author{L.~Kuiper\inst{2} 
\and    W.~Hermsen\inst{2}
\and    G.~Cusumano\inst{5}
\and    R.~Diehl\inst{1}
\and    V.~Sch\"onfelder\inst{1}
\and    A.~Strong\inst{1}
\and    K.~Bennett\inst{3}
\and    M.L.~McConnell\inst{4}
       }

\offprints{L.M.Kuiper$@$sron.nl}

\institute{Max-Planck-Institut f\"ur Extraterrestrische Physik, D-85741 Garching, Germany
  \and     SRON - National Institute for Space Research, Sorbonnelaan 2, NL-3584 CA Utrecht, 
           The Netherlands
  \and     Astrophysics Division, European Space Research and Technology Centre, 2200 AG,
           Noordwijk, The Netherlands
  \and     Space Science Centre, University of New Hampshire, Durham, NH 03824, USA
  \and     Istituto di Fisica Cosmica ed Applicazioni all'Informatica CNR, 
           Via U. La Malfa 153, I-90146, Palermo, Italy
          }

\date{Received  2000 / Accepted 2000}


\abstract{We present the time-averaged characteristics of the Crab pulsar in the 0.75-30 MeV
energy window using data from the imaging Compton Telescope COMPTEL aboard the
Compton Gamma-Ray Observatory (CGRO) collected over its 9 year mission. Exploiting the 
exceptionally long COMPTEL exposure on the Crab allowed us to derive significantly improved
COMPTEL spectra for the Crab nebula and pulsar emissions, and for the first time to accurately 
determine at low-energy $\gamma$-rays the pulse profile as a function of energy. 
These timing data, showing the well-known main pulse and second pulse at a phase 
separation of $\sim 0.4$ with strong bridge emission, are studied together with data 
obtained at soft/hard X-ray energies from the ROSAT HRI, BeppoSAX LECS, MECS 
and PDS, at soft $\gamma$-rays from CGRO BATSE and at high-energy
$\gamma$-rays from CGRO EGRET in order to obtain a coherent high-energy picture 
of the Crab pulsar from 0.1 keV up to 10 GeV. The morphology of the 
pulse profile of the Crab pulsar is continuously changing as a function of energy: 
the intensities of both the second pulse and the bridge emission increase relative 
to that of the first pulse for increasing energies up to $\sim 1$ MeV. 
Over the COMPTEL energy range above 1 MeV an abrupt morphology change happens: the first 
pulse becomes again dominant over the second pulse and the bridge emission loses 
significance such that the pulse profile above 30 MeV is similar to the one observed at optical 
wavelengths. 
A pulse-phase-resolved spectral analysis performed in 7 narrow phase slices consistently 
applied over the 0.1 keV - 10 GeV energy interval shows that the pulsed emission can 
empirically be described with 3 distinct spectral components:
i) a power-law emission component (1 keV - 5 GeV; photon index $2.022\pm 0.014$), 
present in the phase intervals of the two pulses; ii) a curved spectral component 
required to describe soft ($\la 100$ keV) excess emission present in the same pulse-phase 
intervals; iii) a broad curved spectral component reflecting the bridge emission from 0.1 
keV to $\sim 10$ MeV. This broad spectral component extends in phase over the full pulse 
profile in an approximately triangular shape, peaking under the second pulse.
Recent model calculations for a three-dimensional pulsar magnetosphere with outer 
magnetospheric gap acceleration by Cheng et al. (2000) appear at present most successful 
in explaining the above complex high-energy characteristics of the Crab pulsar.  
\keywords{pulsars: individual: \psr -- Stars: neutron -- supernovae: individual: Crab nebula 
-- Gamma rays: observations -- X-rays: stars}}

\maketitle
\markboth{The MeV characteristics of the Crab pulsar}{L.Kuiper et al.}




\section{Introduction}

The Crab pulsar (PSR B0531$+$21) has been studied extensively over the entire electromagnetic spectrum with pulse profiles 
dominated by two pulses, separated $\sim$ 0.4 in pulse phase and approximately aligned in absolute phase over all wavelengths. 

After the first detections of pulsed emission in the X-ray regime by \cite{grcrab_fritz} ; $\sim$ 1-13 keV) and \cite{grcrab_bradt} ; 
1.5-10 keV) , the first significant detection of pulsed soft $\gamma$-ray emission was reported by \cite{grcrab_kurfess}; 100-400 keV).
A great boost forward was made by the X-ray instruments aboard the OSO-8 
(\cite{grcrab_pravdo}; 2-50 keV, revealing spectral variations of the pulsed emission as a function of pulse-phase) , 
HEAO-1 (\cite{grcrab_knight}, 18-200 keV; introducing the possible existence of at 
least 2 pulsed emission components: one associated with the 2 main peaks and one with the bridge, the interval between the
two peaks) 
and HEAO-2 (Einstein) satellites (\cite{grcrab_harnden}, 0.1-4.5 keV; producing the first high-resolution ($\sim 4\arcsec$) 
image of the Crab nebula/pulsar in X-rays along with a high-resolution soft X-ray pulse profile). 
Recently, \cite{grcrab_pravdotwo} presented the details of a pulse-phase-resolved spectral analysis of the pulsed emission 
in the 5-200 keV interval based on RXTE PCA and HEXTE data. They found systematic spectral changes in the photon power-law
index as a function of pulse-phase across the interval of the pulsed emission. Their work is confirmed by the findings
presented by \cite{grcrab_massaro_two}, who used data from the narrow field instruments aboard BeppoSAX (0.1-300 keV). 
These authors also made an attempt to disentangle the two emission components, introduced by Knight (1982), assuming 
for one component (reflecting the emission in the 2 peaks) the shape of the optical pulse profile and for the other component
(the bridge emission) a shape from an analytical model with adjustable parameters. 

At medium $\gamma$-ray energies the first detection of pulsed radiation (0.6-9 MeV) from the Crab pulsar 
was reported by \cite{grcrab_hillier}; see also \cite{grcrab_walraven},
\cite{grcrab_graser}, \cite{grcrab_mahoney} and \cite{grcrab_agrinier}.

In the high-energy $\gamma$-ray domain ($\gtap 30$ MeV) the first indications for pulsed emission from the Crab pulsar 
were obtained from data collected by balloon-borne spark chambers or gas Cherenkov detector systems (see e.g. \cite{grcrab_browning},
\cite{grcrab_albats}, \cite{grcrab_parlier}, \cite{grcrab_mcbreen}). A big step forward in this energy range was made by the SAS-2 spark chamber 
experiment in the early seventies. In these data (20 MeV - 1 GeV) significant pulsed emission was reported by \cite{grcrab_kniffen} and 
\cite{grcrab_thompsonone}.
The most detailed early information on the pulsed high-energy $\gamma$-ray properties of the Crab pulsar was, however, provided 
by the data from the European COS-B satellite (\cite{grcrab_bennett}; \cite{grcrab_wills}; \cite{grcrab_clear}).
Significant bridge emission was discovered in the combined COS-B Crab dataset, and the spectral characteristics of the pulsed
and unpulsed (nebula) emission turned out to be quite diverse. Also, the alignment of the main pulse (P1) from the radio regime up to 
high-energy $\gamma$-rays was shown by Wills et al. (1982). 

The launch of the Compton Gamma-Ray Observatory (CGRO; 20 keV-30 GeV) in April 1991 brought about an enormous improvement in the 
statistical quality of the $\gamma$-ray data. During its exceptionally long lifetime of more than 9 years Crab pulsar data were
collected by the Energetic Gamma-Ray Experiment (EGRET; 20 MeV - 30 GeV) showing clearly changing spectral behaviour as a 
function of pulse-phase (Nolan et al. 1993; Fierro 1995; Fierro et al. 1998). The imaging Compton Telescope COMPTEL (0.75 - 30 MeV) 
viewed the Crab each time simultaneously with EGRET. The results based on data from an early set of observations 
performed during the first-year all-sky survey of the CGRO mission had been published by \cite{grcrab_muchone} and \cite{grcrab_carraminana}. 
\cite{grcrab_ulmer_one} presented the first findings from the Oriented Scintillation Spectrometer Experiment OSSE (50 keV-10 MeV),
and in a CGRO-paper (Ulmer et al. 1995) the first results from OSSE, COMPTEL and EGRET were combined. It was shown
that the overall pulse phase-averaged spectrum is not well fitted by a single power-law, better with a broken power-law. Phase-resolved
spectra were produed and fitted with broken power laws, selecting three phase intervals (peak 1, the bridge and peak 2).

In this paper the final COMPTEL 0.75-30 MeV results on the Crab pulsar/nebula are presented using i) data from 
all available COMPTEL Crab observations, ii) upgraded/improved response estimates and iii) improved data selection criteria. In order to obtain a broad high-energy picture, we also consistently 
analysed in detail Crab data from the Italian/Dutch BeppoSAX satellite at lower energies (0.1--300 keV) and from 
EGRET at higher energies (30 MeV--10 GeV). For some parts of the work we analysed additional data (e.g. soft $\gamma$-ray/X-ray 
data from the CGRO Burst and Transient Source Experiment, BATSE, and the ROSAT HRI; and for comparisons optical and UV data). 


\begin{table}[h]
\caption[]{\label{obs_table} COMPTEL observation summary with \psr \\
 within $30\degr$ from the pointing axis}
\begin{flushleft}
\begin{tabular}{lcccc}
\hline\noalign{\smallskip}
VP \#  & Start Date      & End Date  &Off angle & Exposure        \\
       & TJD$^{\dagger}$ & TJD       &($\degr$) & (1-3 MeV;  \\
       &                 &           &          & $10^6$ cm$^2$s) \\
\hline\noalign{\smallskip}
\multicolumn{5}{l}{Cycle 0} \\
0.3        & 8374.853    & 8377.686    &  8.98        & $\top$   \\
0.4        & 8377.894    & 8380.678    &  8.98        & $8.677$  \\
0.5        & 8380.886    & 8383.662    &  0.13        & $\bot$   \\
\noalign{\smallskip}
\multicolumn{5}{l}{Cycle I} \\
1          & 8392.903    & 8406.785    &  6.51        & $10.725$ \\
31         & 8784.730    & 8798.554    & 27.78        & $ 9.614$ \\
36.0       & 8845.170    & 8846.765    & 15.55        & $\top$   \\
36.5       & 8846.806    & 8854.644    & 16.64        & $12.802$ \\
39         & 8866.263    & 8882.637    & 17.52        & $\bot$   \\
\noalign{\smallskip}
\multicolumn{5}{l}{Cycle II} \\
213        & 9069.778    & 9075.544    &  3.19        & $ 2.764$ \\
221        & 9120.708    & 9131.637    &  3.00        & $ 4.615$ \\
\noalign{\smallskip}
\multicolumn{5}{l}{Cycle III} \\
310        & 9322.653    & 9334.635    & 14.59        & $ 6.387$ \\
321.1/5    & 9391.663    & 9400.636    &  4.49        & $ 7.622$ \\
337        & 9573.925    & 9593.594    & 21.38        & $11.143$ \\
\noalign{\smallskip}
\multicolumn{5}{l}{Cycle IV} \\
412         & 9776.688    & 9783.672    &  6.48        & $ 4.307$ \\
413         & 9783.690    & 9797.589    &  7.54        & $ 9.161$ \\
419.1       & 9811.629    & 9818.586    & 25.92        & $ 4.071$ \\
419.5       & 9846.614    & 9860.634    & 29.17        & $ 6.221$ \\
420         & 9860.654    & 9874.688    & 18.28        & $ 9.462$ \\
426         & 9937.618    & 9951.581    &  0.13        & $ 9.629$ \\
\noalign{\smallskip}
\multicolumn{5}{l}{Cycle V} \\
502         &10007.590    &10021.594    &  8.38        & $10.877$ \\
520         &10210.681    &10224.556    & 24.18        & $ 8.310$ \\
523         &10259.621    &10273.551    & 26.70        & $ 7.462$ \\
526/527/528 &10294.630    &10322.616    &  5.04$^{\ddagger}$ & $20.969$ \\
\noalign{\smallskip}
\multicolumn{5}{l}{Cycle VI} \\
616.1       &10497.670    &10525.647    &  8.56        & $20.046$ \\
\noalign{\smallskip}
\multicolumn{5}{l}{Cycle VII} \\
724.5       &11001.609    &11015.620    &  9.60        & $ 9.852$ \\
\noalign{\smallskip}
\multicolumn{5}{l}{Cycle VIII} \\
816         &11309.621    &11323.581    & 14.59        & $ 8.132$ \\
829         &11435.584    &11449.597    &  3.00        & $10.826$ \\
\noalign{\smallskip}
\multicolumn{5}{l}{Cycle IX} \\
903.1       &11533.662    &11540.639    & 16.78        & $ 4.212$ \\
918.5       &11659.637    &11673.620    &  4.00        & $10.543$ \\
919.5       &11673.644    &11690.999    & 19.76        & $ 9.274$ \\
\noalign{\smallskip}
\hline\noalign{\smallskip}
\multicolumn{5}{l}{$^{\dagger}$\smp\smp TJD = JD - 2440000.5 = MJD - 40000} \\
\multicolumn{5}{l}{$^{\ddagger}$\smp\smp weighted mean of 3 observations } \\
\end{tabular}
\end{flushleft}
\end{table}

\section{Instrument description and observations}

COMPTEL is the imaging Compton Telescope aboard CGRO and operates in the 0.75-30 MeV energy range.
Its detection principle relies on a two layer interaction: a Compton scatter in one of the 7 upper-detector 
(D1) modules followed by a second interaction in one of the 14 lower-detector (D2) modules. 
Main measured quantities are the angles ($\chi,\psi$) specifying the direction of the scattered photon (from 
the interaction loci in D1 and D2) and the energy deposits in the D1/D2 modules where the interactions took place.
From the last two quantities we can calculate the scatter angle $\phibar$ and the total energy
deposit $\etot$ (for a full description see \cite{grcrab_schonfelder}).
Its energy resolution is 5-10\% FWHM and, due to its large field of view of typically 1 
steradian, it is possible to monitor a large part of the sky simultaneously with a position determination 
accuracy of $\sim 1\degr$. The events are time tagged with an accuracy of 0.125 ms and are converted to
Coordinated Universal Time (UTC) with an absolute accuracy better than $100\ \mu$s using the on board
master clock, serving also the other 3 CGRO instruments BATSE, OSSE and EGRET.

In this study we selected all CGRO viewing periods for which the angle between the pointing 
axis (co-aligned with the COMPTEL/EGRET z-axis) and the Crab pulsar is less than $30\degr$.
Details for each individual observation can be found in Table \ref{obs_table}, which is self-explanatory. 

Because we have also included extensively archival EGRET data in the current study, a brief summary of this CGRO instrument
is given as well. EGRET has a (gas-filled) sparkchamber and is sensitive to $\gamma$-rays with energies in the 
range 30 MeV to 30 GeV. In the mode used for most of the observations the field of view is approximately $80\degr$ 
in diameter. Its effective area is approximately $1500\ \hbox{\rm cm}^2$ between 200 and 1000 MeV. 
The angular resolution is strongly energy dependent: the $67\%$ confinement angles at 35 MeV, 500 MeV and 3 GeV 
are $10\fdg 9$, $1\fdg 9$ and $0\fdg 5$ respectively. The energy resolution $\Delta E / E$ is $\sim 20\%$ (FWHM) 
over the central part of the energy range. The relative timing accuracy is $8\ \mu$s and the absolute accuracy 
is better than $100\ \mu$s. 
For a detailed overview of the EGRET detection principle and instrument characteristics, see \cite{grcrab_thompsontwo}. 

The X-ray data most extensively used in this work had been collected with the 4 narrow field instruments aboard 
BeppoSAX: the low-energy (0.1-10 keV) and medium energy (1.6-10 keV) concentrator spectrometers, LECS and MECS
respectively, the High-Pressure Gas Scintillation Proportional Counter, HPGSPC (4-60 keV) and the Phoswich Detector System PDS 
sensitive in the 15-300 keV energy range. Detailed instrument descriptions for the 4 narrow field instruments 
can be found in \cite{grcrab_parmar}, \cite{grcrab_boella}, \cite{grcrab_manzo} and \cite{grcrab_frontera} for the LECS, MECS, HPGSPC 
and PDS respectively.


\section{Timing analysis}

\begin{figure}[t]
              \vspace{-1cm}
              {\hspace{0.5cm}
              \psfig{figure=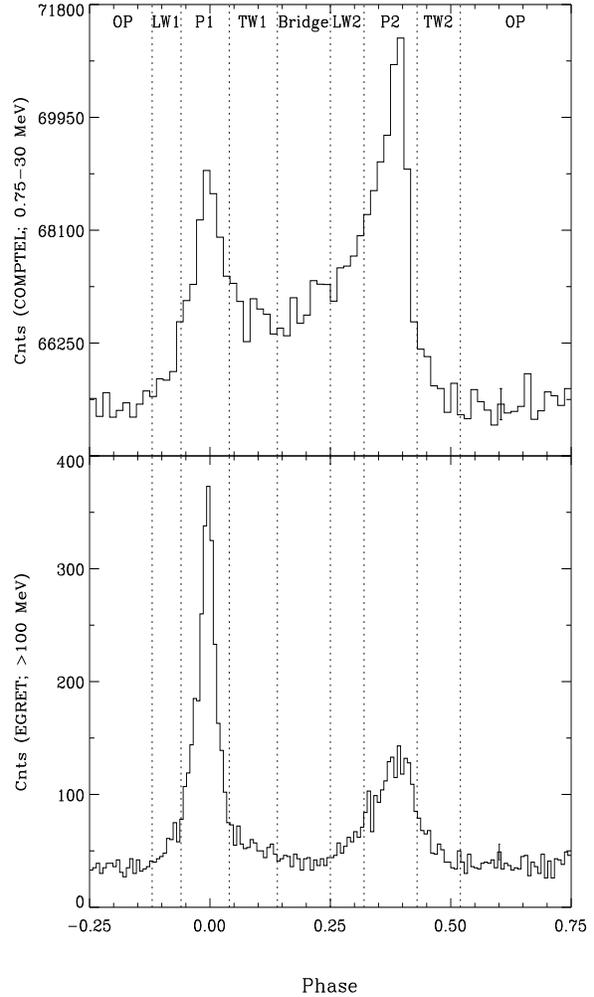,width=8cm,height=15cm}}
              {\caption[]{Pulse profiles of \psr as observed by CGRO COMPTEL in the 0.75-30 MeV
               energy interval (top) and CGRO EGRET (bottom) for energies $>100$ MeV.
               The boundaries of the pulse phase intervals defined by \cite{grcrab_fierro} are
               indicated by the dotted vertical lines. Notice the morphology change: at energies
               between 0.75-30 MeV the second peak dominates, while at energies above 100 MeV
               the first peak dominates. 
               Moreover, in the 0.75-30 MeV interval there is considerable bridge emission in 
               between the two peaks, which is hardly present at energies above 100 MeV.                 
               \label{fig:comptel_egret_lc}}
              }
\end{figure}

The first step in the timing analysis is to subject the events registered during
an observation to an event selection filter. In the case of COMPTEL the most important 
selection parameters are the time-of-flight TOF, the pulse shape discriminator PSD
(see e.g. \cite{grcrab_schonfelder}), the ``spatial'' parameters $\chi,\psi,\phibar$  
and the total energy deposit $\etot$. Given the {\it a priori} known position of the Crab pulsar
it is possible to determine for each event the so-called $\phigeo$ angle, i.e the angle 
between the scattered photon and the source. The $\phibar$ angle
provides an equivalent measure of this angle, but now only based on the energy deposits in
both detector layers. The difference angle $\phiarm= \phibar - \phigeo$ is called the Angular Resolution
Measure (ARM) and forms the base 
of the spatial response of COMPTEL and its distribution is narrowly peaked near $\phiarm= 0$
with asymmetric (energy dependent) wings. The definite and significant timing signature of 
the Crab pulsar in the COMPTEL energy range (Much et al. 1995; \cite{grcrab_muchtwo}) provides a very
usefull tool to determine the optimum event parameter windows for celestial sources. In this
study we have determined and used subsequently the optimum (total energy deposit dependent) 
parameter windows for the TOF and PSD. The optimum windows for the $\phiarm$ angle turn out
to be asymmetric around 0 and a function of total energy deposit (as expected). 
Finally, for the given combination of viewing periods (see
Table \ref{obs_table}) we compared the measured $\phibar$ distribution, dominated by background photons,  
with the distribution expected for a point source at the Crab position. This allows for a 
determination of the optimum window for selection on $\phibar$ (a function of the total energy deposit). 

\begin{table}[t]
\caption[]{\label{tab_pulsecomp} Phase component definitions for the Crab pulsar \\
adopted in this study (see also Fig. \ref{fig:comptel_egret_lc})}
\begin{flushleft}
\begin{tabular}{llcc}
\hline\noalign{\smallskip}
Component  & Abbreviation      & Phase interval  & Width      \\
\hline\noalign{\smallskip}
Leading Wing 1        & LW1    & 0.88 - 0.94     & 0.06   \\
Peak 1                & P1     & 0.94 - 1.04     & 0.10   \\
Trailing Wing 1       & TW1    & 0.04 - 0.14     & 0.10   \\
Bridge                & Bridge & 0.14 - 0.25     & 0.11   \\
Leading Wing 2        & LW2    & 0.25 - 0.32     & 0.07   \\
Peak 2                & P2     & 0.32 - 0.43     & 0.11   \\
Trailing Wing 2       & TW2    & 0.43 - 0.52     & 0.09   \\
\noalign{\smallskip}
Off Pulse             & OP     & 0.52 - 0.88     & 0.36   \\
Total Pulse           & TP     & 0.88 - 1.52     & 0.64   \\
\noalign{\smallskip}
\hline\noalign{\smallskip}
\end{tabular}
\end{flushleft}
\end{table}

\begin{figure}[t]
               \vspace{-1.25cm}
              {\hspace{ 0.00cm}
              \psfig{figure=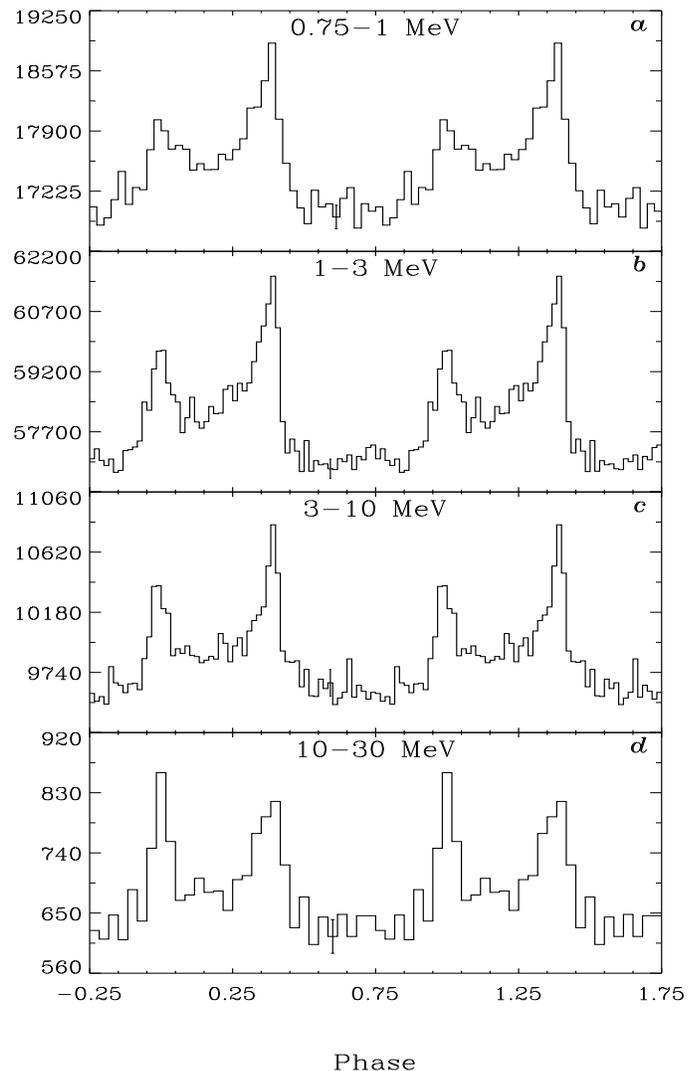,width=9.5cm,height=16cm}}
              {\caption[]{Pulse profiles (double cycles) of \psr as measured 
                by CGRO COMPTEL in 4 differential energy intervals: 0.75-1, 
                1-3, 3-10 and 10-30 MeV. Typical error bars are shown near 
                phase 0.6. A clear morphology change of the pulse profiles 
                is visible: below 10 MeV the second peak (near phase 0.4) 
                dominates, and emission in the ``bridge" phase interval is 
                significant, while above 10 MeV the first peak (near phase 0.0) 
                dominates with strongly reduced ``bridge" emission.
                \label{fig:comptel_lc}}
              }
\end{figure}

Events, not vetoed by any of the 4 anti-coincidence domes surrounding the COMPTEL detector layers, and
having a very low probability of originating from the Earth (``Earth Horizon angle selection'') and 
passing through our optimized event selection windows are finally used in the timing analysis. 

The event arrival times (at the spacecraft) of these accepted events are converted to arrival 
times at the barycentre of the solar system (SSB) using the JPL DE200 solar system ephemeris 
and the Crab pulsar position.
This process yields SSB arrival times with an absolute timing accuracy of less than $100 \mu$s. 
Subsequent phase folding using the Crab pulsar ephemeris data (CGRO timing database; Arzoumanian 1992) yields
the pulse profile. Because the Crab pulsar shows a lot of timing noise (young pulsar) the 
ephemerides have only a limited validity interval. Therefore, in the
phase folding process we used different timing solutions (ephemerides) for (almost) each observation 
given in Table \ref{obs_table}. 

The derived final  -- combining all observations given in Table \ref{obs_table}  --  COMPTEL pulse profile 
over the total energy range 0.75-30 MeV is shown in Fig. \ref{fig:comptel_egret_lc}. The two expected peaks, 
separated $\sim 0.4$ in phase, with intense emission between the peaks (bridge emission) are visible with 
high statistics. The total exposure has increased by a factor $\sim$ 5 compared to the last published total 
COMPTEL profile of the Crab (Much et al. 1995). For comparison we also show in Fig. \ref{fig:comptel_egret_lc} 
the CGRO EGRET profile for energies above 100 MeV for which we analyzed archival Cycle 0 to VI EGRET viewing 
periods with the Crab pulsar within $35\degr$ from the pointing axis, and in which the spark chamber
was switched on. The same ephemerides have been used as for the COMPTEL data. The differences in morphology
between the COMPTEL and EGRET profiles are evident. Following Fierro (1995) we used the EGRET
profile shape to select narrow phase intervals for phase resolved spectroscopy studies (see below). The
intervals are shown in the figure and given in Table \ref{tab_pulsecomp}.
Fig. \ref{fig:comptel_lc} shows the pulse profiles (double cycles for clarity) in the 4 ``standard'' 
COMPTEL energy intervals (0.75-1, 1-3, 3-10 and 10-30 MeV). The significances applying the $Z_{\hbox{\rm \scriptsize n}}^2$-test 
(\cite{grcrab_buccheri}) with 8 harmonics on the unbinned set of pulse-phases are (expressed
in Gaussian sigma's) $20.0\sigma$, $31.7\sigma$, $18.6\sigma$ and $10.9\sigma$, respectively.
Comparing the profiles with those presented in Much et al. (1997) shows the enormous increase in statistical quality, 
especially for energies above 3 MeV, due to the longer exposure in combination with our improved event selection procedures. We see the morphology change from the COMPTEL to the EGRET profile in Fig. \ref{fig:comptel_egret_lc}
occurring over the COMPTEL energy range: below $\sim 10$ MeV the second peak (near phase 0.4) dominates the first peak 
(near 0), and the bridge emission is intense, while above $\sim 10$ MeV the first peak dominates and the bridge emission 
is strongly reduced. 

The $\gamma$-ray profiles in Fig. \ref{fig:comptel_egret_lc} and Fig. \ref{fig:comptel_lc} are time-averaged profiles, 
compiled over many years. Before analysing the profiles further, we first verified the long-term stability of the 
$\gamma$-ray signature (flux and pulse shape).


\section{Long-term variability}

\subsection{Flux variations: total pulsed flux in the 1-10 MeV range}

We studied the time variability of the emission from the Crab pulsar by determining 
the ``pulsed'' flux in differential energy windows as a function of time. 
The ``pulsed'' flux has been derived from the number of excess counts 
in the Total Pulse phase interval (see Table \ref{tab_pulsecomp}) on top of the average emission level in 
the Off Pulse phase interval. For the latter interval we assume that the emission originates 
from the nebula only, although a DC-contribution from the pulsar can not be ruled out.
We show as examples the 1-3 and 3-10 MeV results. These results over the more than 9 year baseline 
(April 1991/ May 2000) are shown in Fig. \ref{fig:timevar}. For the 1-3 MeV energy interval typical 
integration times are 2/3 weeks, while in the 3-10 MeV interval longer integration times are used 
given the strongly reduced statistics. 
The $\chi_{\nu}^2$ values for the fits assuming a constant flux level
are typically $\sim 0.5$ for both energy ranges, indicating that there is no evidence for 
time variability. \cite{grcrab_fierro_two} also studied the Crab long-term variability for energies above 100 MeV
and also concluded that the emission from the Crab pulsar is stable.

{\begin{figure}[t]
  \hspace{-0.1cm}
  \vbox{
  {\psfig{file=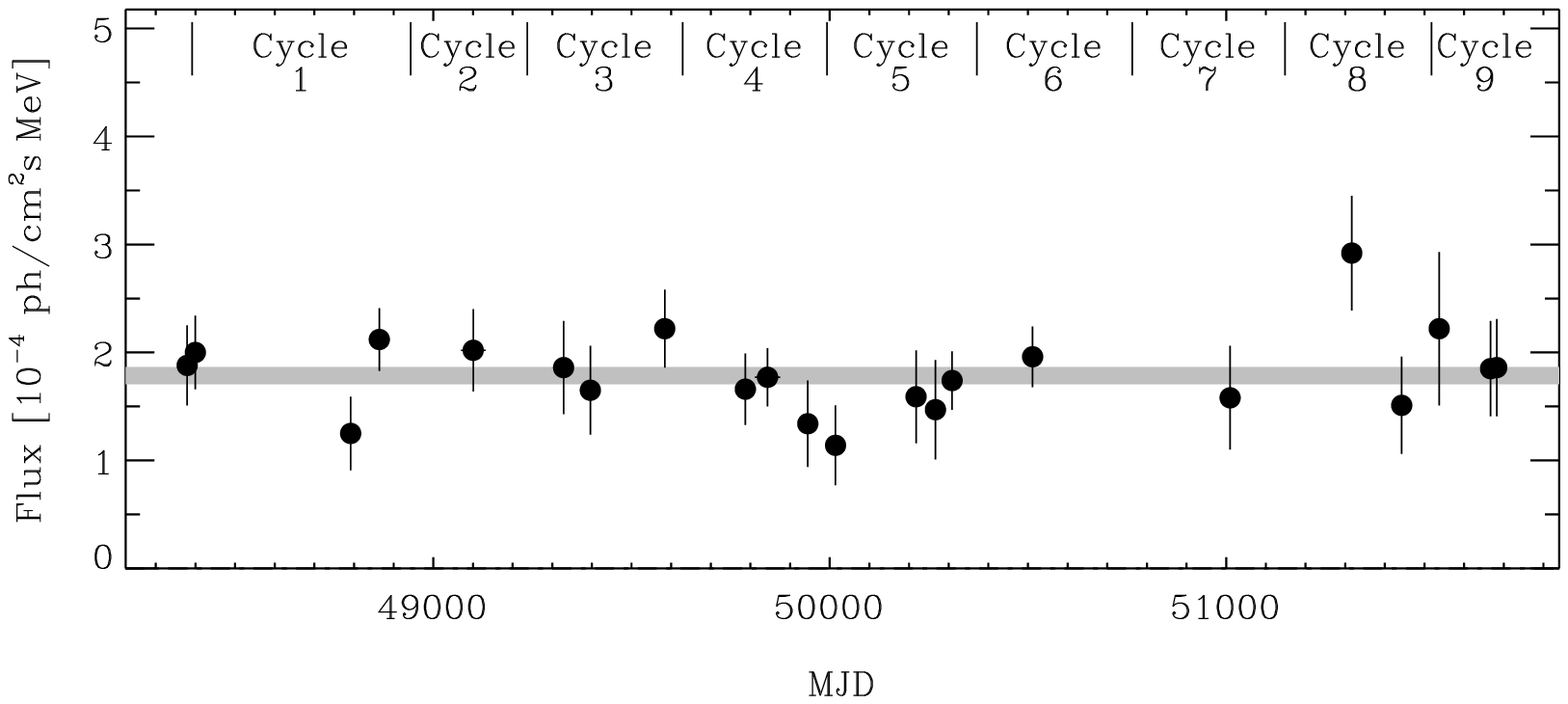,width=9cm,height=5cm}}
  {\psfig{file=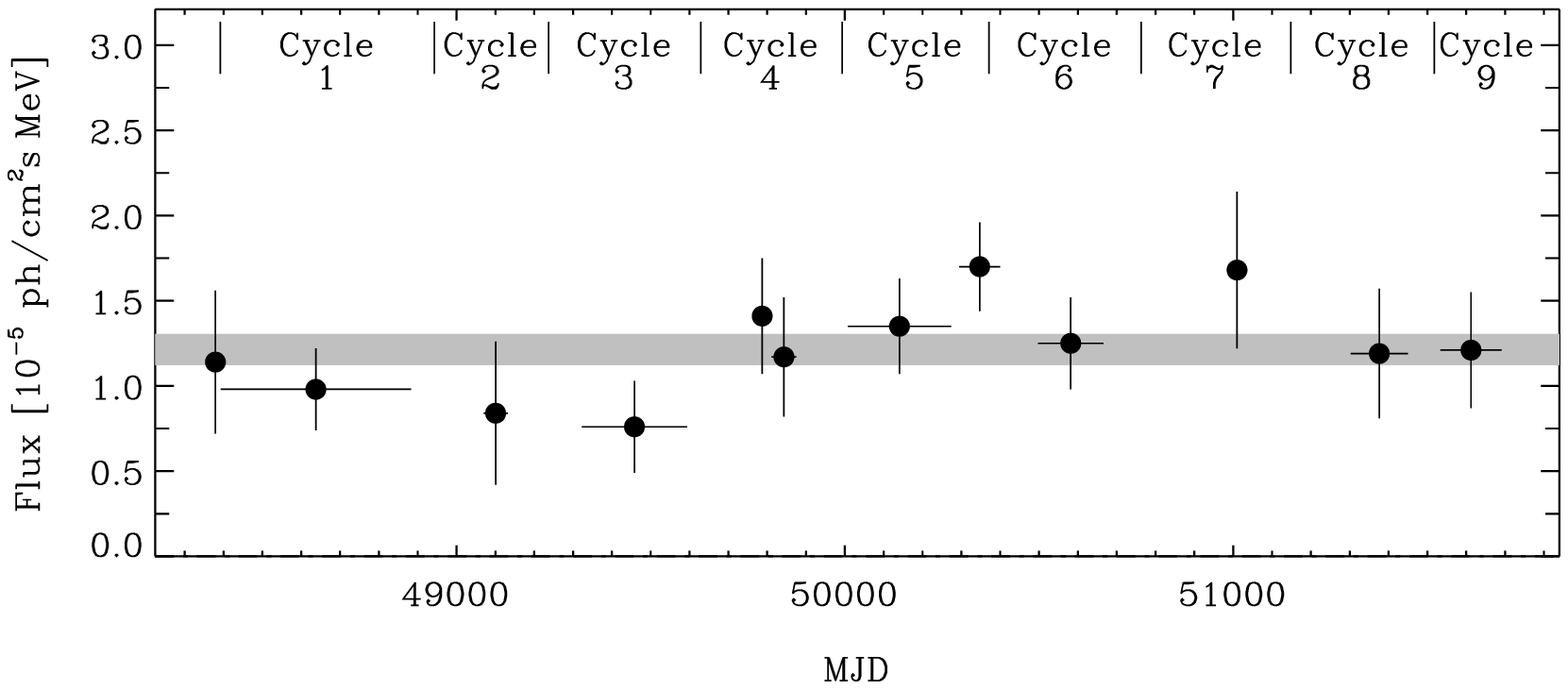,width=9cm,height=5cm}}
       }

  \caption[]{``Total Pulsed'' flux from the 
             Crab pulsar as a function of time in the 1-3 MeV (top) and 3-10 MeV (bottom) energy intervals.
             The $\pm 1\sigma$ uncertainty intervals assuming a constant flux are indicated
             by the shaded regions in both figures. The $\chi_{\nu}^2$-values for the fits
             assuming the flux being constant are $\sim 0.5$, thus there is no indication
             for ``Total Pulsed'' flux variability in both the 1-3 and 3-10 MeV energy intervals.
             \label{fig:timevar}}
\end{figure}}

{\begin{figure}[t]
  \hspace{-0.1cm}
  {\psfig{file=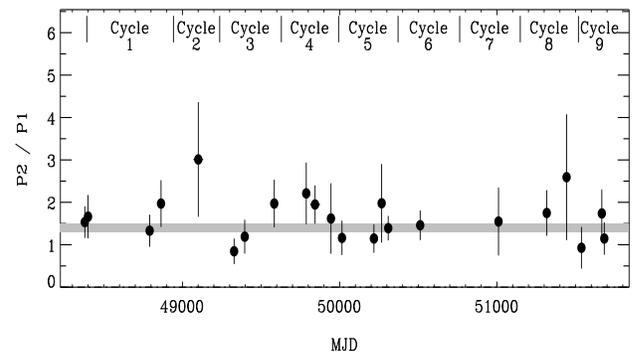,width=9cm,height=5cm}}
  \caption[]{The P2/P1 flux ratio of the Crab pulsar in the 1-3 MeV energy 
             interval as a function of time. The $\pm 1\sigma$ uncertainty interval assuming a constant ratio is indicated
             by the shaded region. The P2/P1 ratio is consistent with being constant.
             \label{fig:peakratio}}
\end{figure}}

\subsection{Pulse shape variations: P2/P1 ratio in the 1-3 MeV range}

The time variability of the pulse shape was investigated in the 1-3 MeV (best statistics) energy window by determining
the P2/P1 ratio for each observation. This ratio is derived by measuring the number of excess counts in the P2 and P1 
phase intervals (see Table \ref{tab_pulsecomp}) on top of the level in the Off Pulse interval. This ratio is shown in 
Fig. \ref{fig:peakratio}. The $\chi_{\nu}^2$ value for a fit assuming a constant P2/P1 ratio is $\sim 0.8$. 
Therefore, there is no indication for a time dependency of the pulse shape at medium energy $\gamma$-ray
energies consistent with the findings presented by \cite{grcrab_carraminana}.
For energies above 30 MeV \cite{grcrab_fierro_two} studied the long-term temporal variation of the P2/P1 ratio and found
also no evidence for a (systematic) variation over the Cycle 0-III EGRET observations. \cite{grcrab_tompkins} came to
a similar conclusion using an extended EGRET data base including also Cycle IV-V. 


\section{Pulse profiles of \psr from 0.1 keV up to 10 GeV}

The $\gamma$-ray pulse profiles in Fig. \ref{fig:comptel_lc} show that the pulse morphology changes 
significantly over the COMPTEL energy window (0.75-30 MeV), i.e. the emission spectra vary significantly 
with phase. Phase-resolved spectral analyses have earlier been performed at X-ray and $\gamma$-ray energies 
for different data sets and/or different narrow energy intervals. However,
for each study, different phase selections have been made such that a consistent full 
high-energy picture of the Crab pulsar can not be compiled from published results. Therefore, we extended 
our energy window by analysing {\em consistently} not only the CGRO EGRET (30 MeV - 10 GeV) high-energy 
$\gamma$-ray data, but also X-ray/soft $\gamma$-ray data from the ROSAT HRI (0.1-2.4 keV), BeppoSAX LECS 
(0.1-10 keV), MECS (1.6-10 keV) and PDS (15-300 keV), and CGRO BATSE (20 keV- 1 MeV). 

The (on board folded) data from CGRO BATSE overlap in energy with the data from CGRO OSSE for which results have 
already been published by Ulmer et al. (1994,1995). However, due to the enormous exposure in the co-added BATSE 
data the statistics are much better than can be obtained in the combined OSSE Crab observations.
Especially above $\sim 220$ keV (e.g. Ulmer et al. 1994, Fig. 2) where the OSSE data have low statistical quality, 
the BATSE profiles are superior.
For the high statistics OSSE data (below $\sim$ 220 keV) we verified that the profiles are
consistent in shape with those obtained by us using BeppoSAX PDS and CGRO BATSE data.

In the next subsections details are given about the compilation of Crab pulse profiles over the energy range 0.1 keV 
to 10 GeV.

\subsection{ROSAT HRI 0.1-2.4 keV pulse profile}

\begin{figure}[t]
              {\vspace{-1.25cm}}
              {\hspace{-0.25cm}
              \psfig{figure=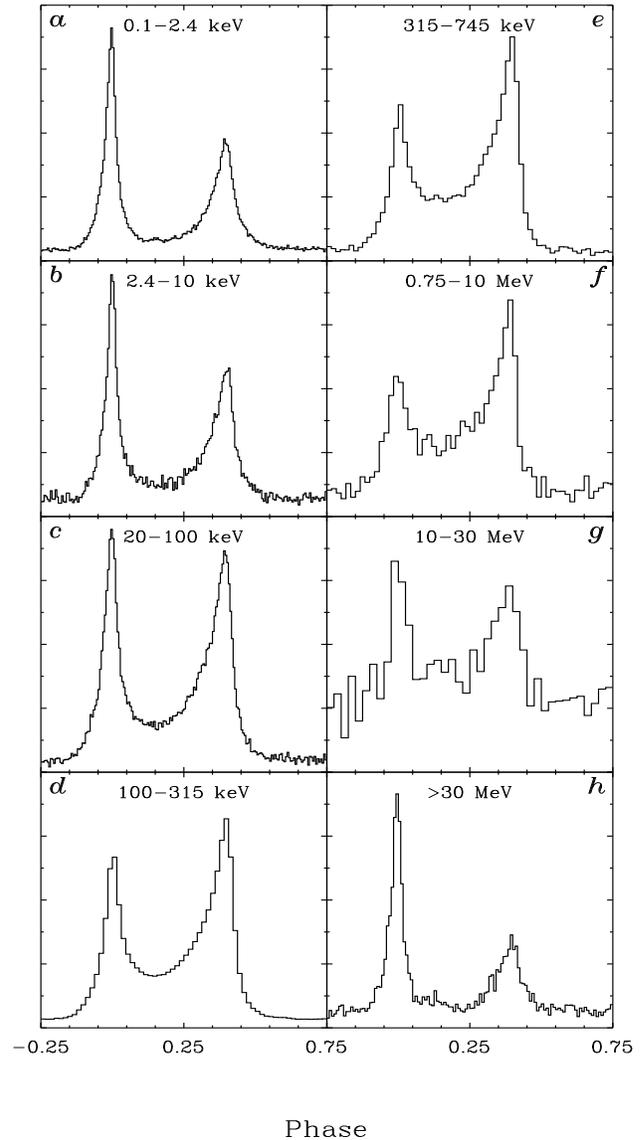,width=9.5cm,height=17cm}}
              {\caption[]{High-energy pulse profiles of \psr from 0.1 keV up to
               10 GeV. Data have been used from the following instruments:
               {\bf a}) ROSAT HRI (0.1-2.4 keV), {\bf b}) BeppoSAX MECS (2.4-10 keV)
               , {\bf c}) BeppoSAX PDS (20-100 keV), {\bf d,e}) CGRO BATSE 
               (100-315 keV \& 315-750 keV), {\bf f,g}) CGRO COMPTEL
               (0.75-10 MeV \& 10-30 MeV) and {\bf h}) CGRO EGRET ($>30$
               MeV). The morphology change of the profiles as a
               function of energy is striking.             
               \label{fig:he_profiles}}
              }
\end{figure}

The soft X-ray ROSAT HRI data were collected during an observation of the Crab pulsar/nebula
performed from 4 March 1995 to 15 March 1995 yielding a net exposure time of 7.98 ks 
({\it HEASARC} Online Service; observation identifier {\it RH400639N00}).
Because the data are spread over 115 different orbital intervals over the 11 day observation 
period the considerable ROSAT clock drift will result in a messy pattern when combining the pulse 
phases from the entire observation. We could identify 4 consecutive sets of orbital intervals
in which the observed pulse profile is stable. The 0.1-2.4 keV pulse profile shown in 
Fig. \ref{fig:he_profiles}{\bf{a}} was obtained cross-correlating 3 of the 4 profiles with the
profile chosen as template, correcting for the observed mutual phase shifts and 
fixing the zero phase at the centre of the main peak.

\subsection{BeppoSAX 2.4-100 keV pulse profiles}

The BeppoSAX LECS, MECS and PDS data have been collected during a calibration observation 
of the Narrow Field instruments aboard BeppoSAX performed on 25-26 September 1999 yielding
(screened) effective exposure times of 7.75 ks, 32.6 ks and 30.7 ks for the LECS, MECS (unit-2)
and PDS clusters A \& B, respectively (data retrieved from archive maintained by BeppoSAX ASI
Science Data Center at {\it http://www.asdc.asi.it/bepposax/}; Observation Codes 20795007 \&
207950071).  In Fig. \ref{fig:he_profiles}{\bf{b}} and {\bf{c}} the
Crab pulse profiles are shown as observed by the MECS in the 2.4-10 keV energy window 
and by the PDS in the 20-100 keV energy window, respectively.

\subsection{CGRO BATSE 100-745 keV pulse profiles}

In the hard X-ray/soft $\gamma$-ray band (0.05-1 MeV) we have used archival data from the CGRO BATSE Large Area 
Detectors collected during observations performed between MJD 48392 and 50273 in the onboard folding 
mode (CGRO Archive maintained by {\it HEASARC} at {\it ftp://cossc.gsfc.nasa.gov/compton/data} directory {\it batse/pulsar/onboard\_folded/crab/}).
 Typical integration times were 2/3 weeks per included observation (64 observations have been 
used in this study). The profiles had been produced in 64 bins per cycle in 16 different energy 
channels for each individual observation run. 
We determined the shifts of the pulse profiles of the individual observation runs with respect to the 
profile obtained during the observation run 48392-48406 in channel 9 ($\sim165$-$230$ keV) by 
cross-correlation. Applying the shifts in the combination of the pulse profiles and putting the first 
peak at phase 0 yields high quality pulse profiles in the 20 keV - 2 MeV range. In Fig. \ref{fig:he_profiles}{\bf{d}} 
and {\bf{e}} the profiles are shown for the 100-315 keV (channels 7-10) and 315-745 keV 
(channels 11-13) energy windows, respectively.

\subsection{CGRO COMPTEL/EGRET pulse profiles}

At medium energy $\gamma$-rays the CGRO COMPTEL pulse profiles derived in this work are shown for the 
0.75-10 and 10-30 MeV energy windows in Fig. \ref{fig:he_profiles}{\bf{f}} and {\bf{g}}, respectively.
Note that the COMPTEL profiles have a large non-zero offset.
Finally, in Fig. \ref{fig:he_profiles}{\bf{h}} the CGRO EGRET pulse profile is given for energies above
30 MeV (we used data from Cycle 0-VI observations, retrieved from the CGRO Archive 
maintained by {\it HEASARC}; {\it ftp://cossc.gsfc.nasa.gov/compton/data/egret/high\_level/}).

\section{P2/P1 and Bridge/P1 ratios as a function of energy}

From the (high-energy) pulse profile compilation shown in Fig. \ref{fig:he_profiles} we can immediately
observe some striking features. The second peak (near phase 0.4) becomes more and 
more pronounced for increasing energies. However, above $\sim$ 10 MeV the first peak
becomes dominant again. The ``Bridge'' emission seems to show a similar behaviour as the second peak.
In a more quantitative evaluation of this morphology change of the profile as a function of energy 
we determined the intensity ratios for P2/P1
and Bridge/P1 as a function of energy over the entire range 0.1 keV to 10 GeV, adopting the
phase interval definitions of Table \ref{tab_pulsecomp}. The pulsed emission in each interval has been separated from 
the underlying nebula/DC emission by subtracting the (properly scaled) emission from the OP phase interval.
The results are visualized in Figs. \ref{fig:peak_ratio} \& \ref{fig:bridge_peak_ratio} for the P2/P1 and Bridge/P1
ratios, respectively. In these plots we have also included the ratios derived from the optical profile in the 3800 to
6500 \AA \ wavelength interval obtained by \cite{grcrab_muchthree} using the UCL MIC detector as well as those 
in the far-ultraviolet (1140-1720 \AA) and near-ultraviolet (1600-3200 \AA), obtained from (time-tagged) 
data taken by the HST STIS instrument (Sollerman et al. 2000, Gull et al. 1998). In all optical ranges we again 
applied the consistent phase interval definitions.

{\begin{figure}[t]
  \hspace{0.3cm}
  {\psfig{file=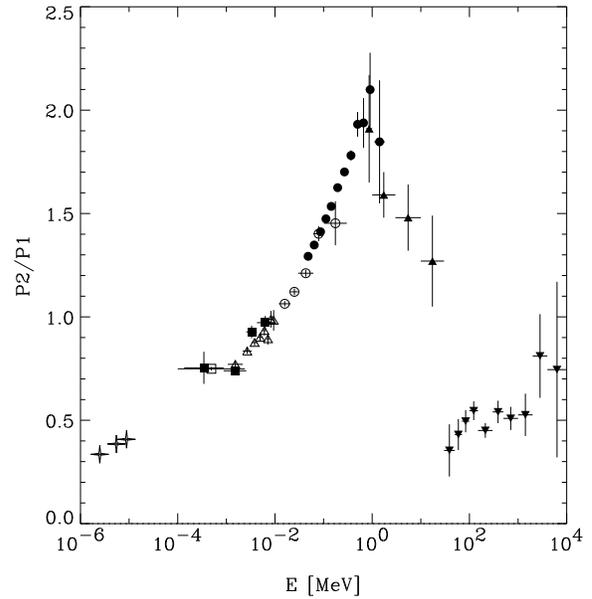,width=8cm,height=8cm}}
  \caption[]{P2/P1 ratio as a function of energy from optical wavelengths up to
             high-energy $\gamma$-rays. Data from the following instruments have
             been used: optical wavelengths, UCL MIC detector (star symbol), NUV/FUV HST 
             STIS (star symbols); X-ray energies, ROSAT HRI (open square), BSAX LECS 
             (filled square), BSAX
             MECS (open triangle); Hard X-rays/soft $\gamma$-rays, BSAX PDS 
             (open circles), CGRO BATSE (filled circles); Medium/hard $\gamma$-ray
             energies, CGRO COMPTEL (filled upwards pointing triangle), CGRO EGRET
             (filled downwards pointing triangle). The gradual increase of the P2/P1
             ratio up to $\sim 1$ MeV is striking, a sharp decline in the
             1-30 MeV energy range follows and a recovery to the optical ratio value
             settles above $\sim$ 30 MeV.
             \label{fig:peak_ratio}}
\end{figure}}

The P2/P1 ratio as a function of energy (Fig. \ref{fig:peak_ratio}) gradually increases from the optical wavelength 
range to $\sim 1$ MeV, followed by a rapid decrease in the 1-30 MeV interval (the COMPTEL energy window) towards 
a more or less constant value of $\sim 0.5$ for energies above 30 MeV (the EGRET energy window). The Bridge/P1 ratio 
vs. energy (Fig. \ref{fig:bridge_peak_ratio}) exhibits a very similar shape as for the P2/P1 ratio.  However, the 
values of the latter ratio in the optical and high-energy $\gamma$-ray domains, become very small ($0.017$) 
indicating that the Bridge emission practically vanishes.  
It is only substantial in the $\sim 1$ keV to $\sim 10$ MeV energy window in contrast with the emissions
from the 2 peaks which are always present. This behaviour suggests that we are dealing with an emission 
component distinct from the emission from both peaks. This hypothesis (see e.g. also \cite{grcrab_knight}; Hasinger 
1984,1985; Massaro et al. 2000) is further strengthened in the phase-resolved spectral analysis presented in the 
next section. 

{\begin{figure}[t]
  \hspace{0.3cm}
  {\psfig{file=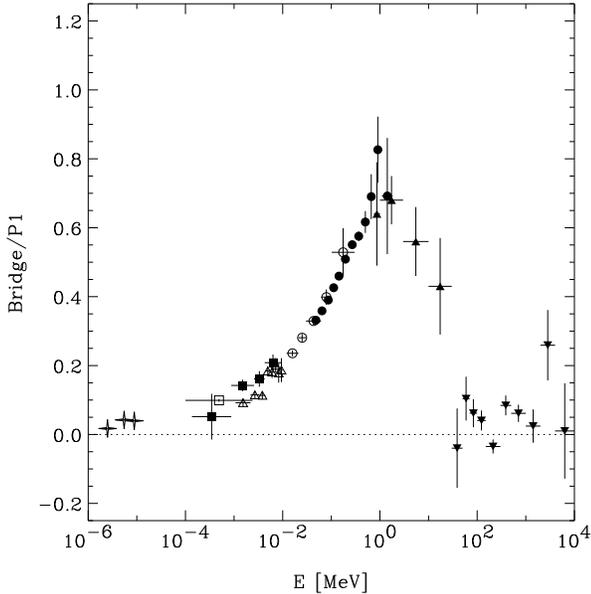,width=8cm,height=8cm}}
  \caption[]{Bridge/P1 ratio as a function of energy from optical wavelengths up to
             high-energy $\gamma$-rays. See the caption of Fig. \ref{fig:peak_ratio}
             for the meaning of the symbols. The Bridge/P1 ratio is almost 0 at
             optical wavelengths, but gradually reaches a maximum near 1 MeV, followed
             by a drastic break in the 1-30 MeV energy range. Above $\sim 30$ MeV
             the Bridge/P1 ratio approaches the optical value of $\sim 0.017$.
             \label{fig:bridge_peak_ratio}}
\end{figure}}

Similar analyses have been presented in the past by other authors (e.g. Toor \& Seward 1977; Hasinger 1984,1985; 
Ulmer et al. 1994; Mineo et al. 1997; Massaro et al. 1998,2000), generally over more restricted energy windows 
with poorer data coverage, and/or often using data of inferior statistical quality.
Particularly the ``transition" region of the COMPTEL MeV window is now well covered for the first time.


\section{Spectral analysis}
\label{section:crab_spa}
In this section we will first present the spectra of the nebula emission
and the Total Pulse emission (excess emission in Total Pulse interval; see Table \ref{tab_pulsecomp}).
Then we will show the results from the phase-resolved spectral analysis for the narrow phase intervals.
As for the timing analysis, we did not limit ourselves to the analysis of the COMPTEL data, but we 
collected data over a very wide spectral band to derive a consistent overall high-energy picture of the Crab. 
Combining spectra derived by different instruments, we had to assess possible systematic effects in the flux
estimates and their impact on our analysis and conclusions. For this purpose, we used the nebula spectra to 
estimate possible inconsistencies.


\subsection{Crab nebula spectrum}
\label{section_opspec}
To determine the Crab nebula spectrum from our COMPTEL (0.75-30 MeV) Cycle 0-IX observations (see Table \ref{obs_table})  
we selected events recorded in the Off Pulse phase interval (see Table \ref{tab_pulsecomp}), assuming that any pulsar DC
emission is negligible, and applied a maximum likelihood method using the spatial signature of a point source in the 
($\phibar,\phiarm$) plane as a function of measured energy. 
This work yielded an improved spectrum for the nebula emission in the COMPTEL energy range compared to the 
COMPTEL Crab nebula spectrum published earlier by van der Meulen et al. (1998). In the latter work a smaller database was used, 
as well as preliminary response characteristics, which have since been improved upon. The newly derived COMPTEL nebula
spectrum is given in Table \ref{cmpflux_table}. A power-law fit to the COMPTEL nebula flux points results in a photon index of 
$2.227\pm0.013$. In the same table also the statistical uncertainties ($1\sigma$) on the flux
measurements are provided. In this context we note that COMPTEL has an overall (systematic) uncertainty on its flux 
estimates of the order of $10-20\%$. 

To cover the neighbouring soft X-ray to soft $\gamma$-ray band, we derived the BeppoSAX LECS, MECS, HPGSPC and 
PDS nebula spectra applying the most recent (December 1999 issue for the LECS and HPGSPC and November 1998 issue for the 
MECS and PDS) response characteristics (sensitive area, energy redistribution matrices and spatial response).
We fitted the Sept. 1999 Off Pulse Crab data from these four BeppoSAX instruments simultaneously over the full 0.1-300 keV energy range 
with an absorbed power-law model taking into account the mutual uncertainties in absolute flux calibrations by including 
in the fit three free relative normalization scale factors (MECS scale factor fixed to 1). The energy of the LECS events used in 
the fit was constrained to the 0.1-4 keV window.
The best photon index and Hydrogen column density N$_{\hbox{\rm \scriptsize H}}$, assuming solar abundances for the other 
elements absorbing soft X-rays, resulting from this fit are $2.145 \pm 0.001$ and $3.61(2)\times 10^{21}\ \hbox{\rm cm}^{-2}$, 
respectively. In Appendix \ref{app_flxcal} the systematic uncertainties in the derived spectral characteristics in the X-ray/soft
$\gamma$-ray band are discussed in detail comparing flux estimates obtained by different high-energy instruments. 

\begin{table}[t]
\caption[]{\label{cmpflux_table} COMPTEL spectra of the Crab nebula and pulsar (Total Pulse).
Fluxes with $1\sigma$ statistical uncertainties}
\begin{flushleft}
\begin{tabular}{rlcc}
\hline\noalign{\smallskip}
\multicolumn{2}{c}{Energy} & Nebula            & Total Pulse           \\
\multicolumn{2}{c}{window} & Flux              & Flux                  \\
\multicolumn{2}{c}{[MeV]}  & [ph/cm$^2$s MeV] & [ph/cm$^2$s MeV]     \\
\hline\noalign{\medskip}
0.75       & 1.00        & $(2.585 \pm 0.089)$E-3 & $(0.650 \pm 0.071)$E-3 \\
1.00       & 1.25        & $(1.563 \pm 0.054)$E-3 & $(0.452 \pm 0.043)$E-3 \\
1.25       & 1.50        & $(1.127 \pm 0.043)$E-3 & $(0.270 \pm 0.034)$E-3 \\
1.50       & 2.00        & $(0.617 \pm 0.020)$E-3 & $(0.165 \pm 0.016)$E-3 \\
2.00       & 2.50        & $(0.306 \pm 0.014)$E-3 & $(0.084 \pm 0.011)$E-3 \\
2.50       & 3.00        & $(0.217 \pm 0.010)$E-3 & $(0.048 \pm 0.008)$E-3 \\
3.00       & 4.00        & $(1.312 \pm 0.055)$E-4 & $(0.278 \pm 0.045)$E-4 \\
4.00       & 6.00        & $(0.613 \pm 0.022)$E-4 & $(0.126 \pm 0.018)$E-4 \\
6.00       & 8.00        & $(0.284 \pm 0.014)$E-4 & $(0.062 \pm 0.011)$E-4 \\
8.00       & 10.0        & $(1.637 \pm 0.082)$E-5 & $(0.288 \pm 0.066)$E-5 \\
10.0       & 15.0        & $(0.734 \pm 0.033)$E-5 & $(0.244 \pm 0.027)$E-5 \\
15.0       & 30.0        & $(0.201 \pm 0.013)$E-5 & $(0.039 \pm 0.011)$E-5 \\
\noalign{\medskip}
\hline\noalign{\medskip}
\end{tabular}
\end{flushleft}
\end{table}

For the adjacent EGRET high-energy (30-30000 MeV) $\gamma$-ray range we derived (phase-resolved) Crab spectra using
all available Cycle 0-IV archival EGRET data for which reliable sensitivity estimates were available to us\footnote{Due 
to the problem of gas aging, the spark chamber efficiency degraded significantly during the later Cycles of the 
EGRET CGRO observations. Recently, the EGRET team did present the energy and time dependent correction factors for the later Cycles
(Gamma 2001 Symposium, Baltimore).}.
The method is equivalent to the spatial maximum likelihood analysis performed by Fierro (1995). 
The only difference with the latter work is that we now added Cycle-IV data.
The EGRET spectral data are claimed to be $10\%$ accurate (\cite{grcrab_thompsonfour}).


{\begin{figure*}[t]
  {\psfig{file=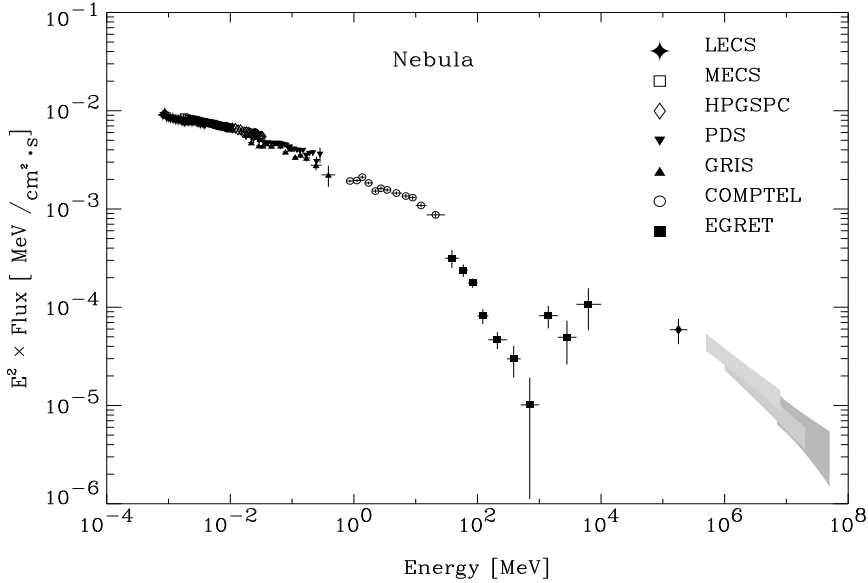,width=12cm,height=8cm}}
  \hfill
  \parbox[b]{55mm}{\vspace{-8cm}
  \caption[]{The Crab nebula spectrum from soft X-rays up to TeV $\gamma$-rays.
             The TeV data point near $1.6\times 10^5$ MeV is taken from Oser et al. (2001) 
             and the hatched bands represent the flux measurements and corresponding $1\sigma$ uncertainty 
             estimates at TeV energies (for references, see text).\vspace{1truecm}
             \label{fig:op_spectrum}}}
\end{figure*}}

{\begin{figure*}[t]
  {\psfig{file=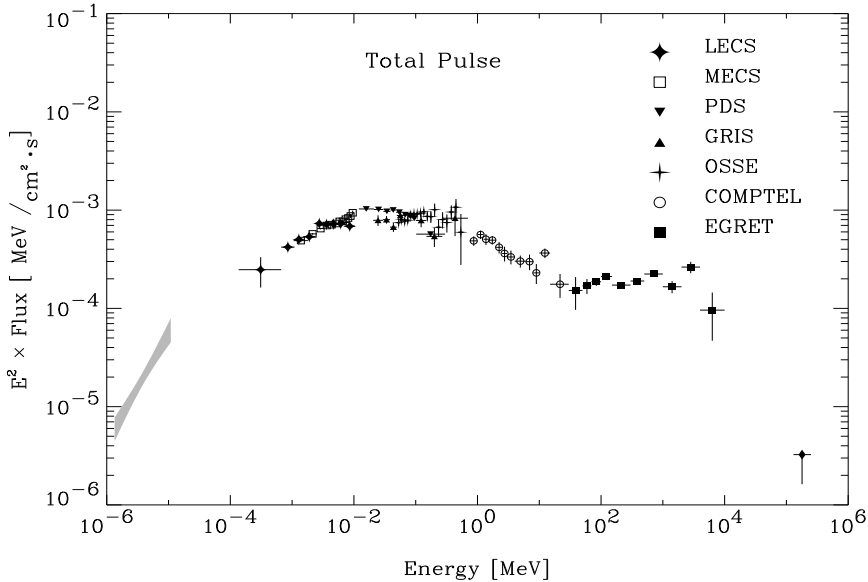,width=12cm,height=8cm}}
  \hfill
  \parbox[b]{55mm}{\vspace{-8cm} 
  \caption[]{The Total Pulse emission of the Crab pulsar from optical wavelengths up to high-energy
             $\gamma$-rays. The nebula emission has been subtracted. The optical spectral data ($10^{-6}$-$10^{-5}$ MeV) 
             are taken from Sollerman et al. (2000) and the TeV data point near $1.6\times 10^5$ MeV from Oser et al. (2001).
             \vspace{1truecm} 
             \label{fig:tp_spectrum}}}
\end{figure*}}


The Crab nebula spectrum from 0.1 keV up to 50 TeV is shown in an $E^2\times F$ representation in Fig. \ref{fig:op_spectrum}.
Included are the BeppoSAX, COMPTEL and EGRET spectra derived in this work, soft $\gamma$-ray spectral information from GRIS
(0.02-1 MeV; Bartlett et al. 1994a) together with ground-based TeV data (STACEE-32 $>0.19$ TeV, \cite{grcrab_oser}; HEGRA 1-20 TeV, 
\cite{grcrab_aharonian}; Whipple 0.5-8 TeV, \cite{grcrab_hillas}; CANGAROO 7-50 TeV, \cite{grcrab_tanimori}).
Similarly to the spectrum shown in van der Meulen et al. (1998), but now more pronounced, we see a continuous and smooth decrease 
from soft X-rays up to medium energy $\gamma$-rays, irrespective of the uncertainties in the absolute sensitivities of the instruments, followed by a steep gradient beyond $\sim 30$ MeV to $\sim 300$ MeV.
Above $\sim 300$ MeV an additional emission feature seems to emerge reaching a maximum between 10 and 100 GeV, a window which is
not yet accessible for space-borne and ground-based experiments. For an interpretation of this spectral shape, see e.g. \cite{grcrab_dejager}.

\subsection{Crab Total Pulse spectrum}
\label{section_tpspec}
In the 0.75-30 MeV energy range we determined the pulsed flux values from all CGRO COMPTEL  
observations (see Table \ref{obs_table}) in two distinct manners. In the first method, the number of (pulsed) excess 
counts in the broad Total Pulse interval (see Table \ref{tab_pulsecomp}) above the mean level in the 
Off Pulse interval is determined as a function of measured energy. These excess counts are then converted to flux 
values using the COMPTEL response and the total exposure.
The second approach is based on the maximum likelihood method in the spatial domain as 
introduced in Sect. \ref{section_opspec}. Applying the latter approach for pulse-phase selected events and 
subtracting the properly scaled Off Pulse contribution (containing only the DC/nebula source with a point source signature) 
yields the pulsed fluxes as a function of energy for the selected pulse-phase intervals. 
We verified for the Total Pulse interval, having the best statistics, that the fluxes derived from both methods are 
compatible within $5-20\%$, giving a measure of the systematic uncertainties. The COMPTEL Total Pulse fluxes from the 
spatial analysis (remember, nebula emission subtracted) are also included in Table \ref{cmpflux_table} and shown in Fig. 
\ref{fig:tp_spectrum}. 

For comparison, we included in Fig. \ref{fig:tp_spectrum} our derived Total Pulse Crab EGRET spectrum  
(30 MeV-10 GeV, Cycle 0-IV) and the published Total Pulse spectra from GRIS (20 keV - 1 MeV: Bartlett 1994a) and OSSE 
(50 keV - 0.59 MeV: Ulmer et al. 1994). In the latter two publications slightly different phase intervals have been used to derive the Total Pulse spectrum.

In the COMPTEL Total Pulse spectrum a feature becomes apparent: the high flux value in the 10-15 MeV interval, consistently derived 
in both the timing and spatial methods. A response anomaly is excluded, e.g. the nebula 0.75-30 MeV 
spectrum (cf. Fig. \ref{fig:op_spectrum}) exhibits a very smooth behaviour over its entire range, nor has such an effect been seen
in other COMPTEL analyses. Fitting the COMPTEL flux points with a power-law spectral model, excluding the deviant 10-15 MeV flux 
point, yields a good fit with photon index of $2.35\pm 0.06$ ($\chi_{\nu}^2=0.50$ for 9 d.o.f.), connecting smoothly at both ends 
to the GRIS/OSSE and EGRET flux measurements. 
The excess flux in the 10-15 MeV interval above this power-law model fit reaches a significance of $3.5\sigma$. Including the deviant 
10-15 MeV flux point in the spectral fit yields a worse fit ($\chi_{\nu}^2=1.42$ for 10 d.o.f) with a power-law index of 
$2.24\pm 0.04$, which does not connect smoothly to the neighbouring measurements, particularly to EGRET. 
The 10-15 MeV excess flux above this fit has a significance of only $2.0\sigma$. We do therefore not regard the 10-15 MeV flux enhancement 
as a firm detection of a new spectral feature, and have no possible astrophysical interpretation, but we find it interesting to note 
that contributions to this flux enhancement appear to come from those (narrow) phase intervals in which a broad spectral 
(Bridge) component dominates the spectrum (see Sect. \ref{section_narspc}). 
If genuine, it could therefore be related to this spectral component. 

For the BeppoSAX LECS, MECS and PDS we determined the number of (pulsed) excess counts above 
the mean level in the Off Pulse phase interval, similarly to the first COMPTEL method. 
These excess counts have been converted to flux measures applying the most recent response characteristics 
assuming intrinsic power-law type emission absorbed in a column of density 
N$_{\hbox{\rm \scriptsize H}}=3.61(2)\times 10^{21}\ \hbox{\rm cm}^{-2}$ (see Sect.\ref{section_opspec}).
In this way we obtained the Total Pulse spectrum over the 0.1-300 keV energy interval, which is also shown in Fig. \ref{fig:tp_spectrum}. 

We augmented the energy coverage by including the pulsed spectra obtained at 
optical/NUV/FUV wavelengths by Sollerman et al. (2000). 
In this $E^2 \times F$ representation, the total pulsed emission shows a gradual increase 
from the optical range towards a plateau of maximum luminosity extending from $\sim 10$ keV to
$\sim 1$ MeV. Beyond $\sim 1$ MeV the emission softens until $\sim 70$ MeV, above which a second plateau appears with
an emission spectrum having a photon power-law index close to $2$. Between 4 and 10 GeV the pulsar spectrum
appears to break/soften drastically to account for the recently derived upper limits for pulsed emission at TeV 
$\gamma$-ray energies (see e.g. \cite{grcrab_vacanti}, \cite{grcrab_borione}, \cite{grcrab_aharonian_two}, 
\cite{grcrab_lessard} and \cite{grcrab_oser}).
Its spectral behaviour is completely different from that of the nebula (cf. Fig. \ref{fig:op_spectrum}).  
Notice the dominance of the nebula emission component over the pulsed emission component for energies below $\sim 100$ MeV
and above $\sim 10$ GeV, comparing Figs. \ref{fig:op_spectrum} \& \ref{fig:tp_spectrum}. Only in a small window at 
high-energy $\gamma$-rays between $\sim 100$ MeV and $\sim 10$ GeV the pulsed component exceeds the underlying nebula component.
This Total Pulse spectrum is clearly complex. For detailed theoretical interpretations it is important to disentangle first
this total spectrum in contributions from different phase components of the pulse profile. 


{\begin{figure*}[t]
  \hspace{0.3cm}
  \vbox{
        \hbox{\hspace{-0.5cm}  {\psfig{file=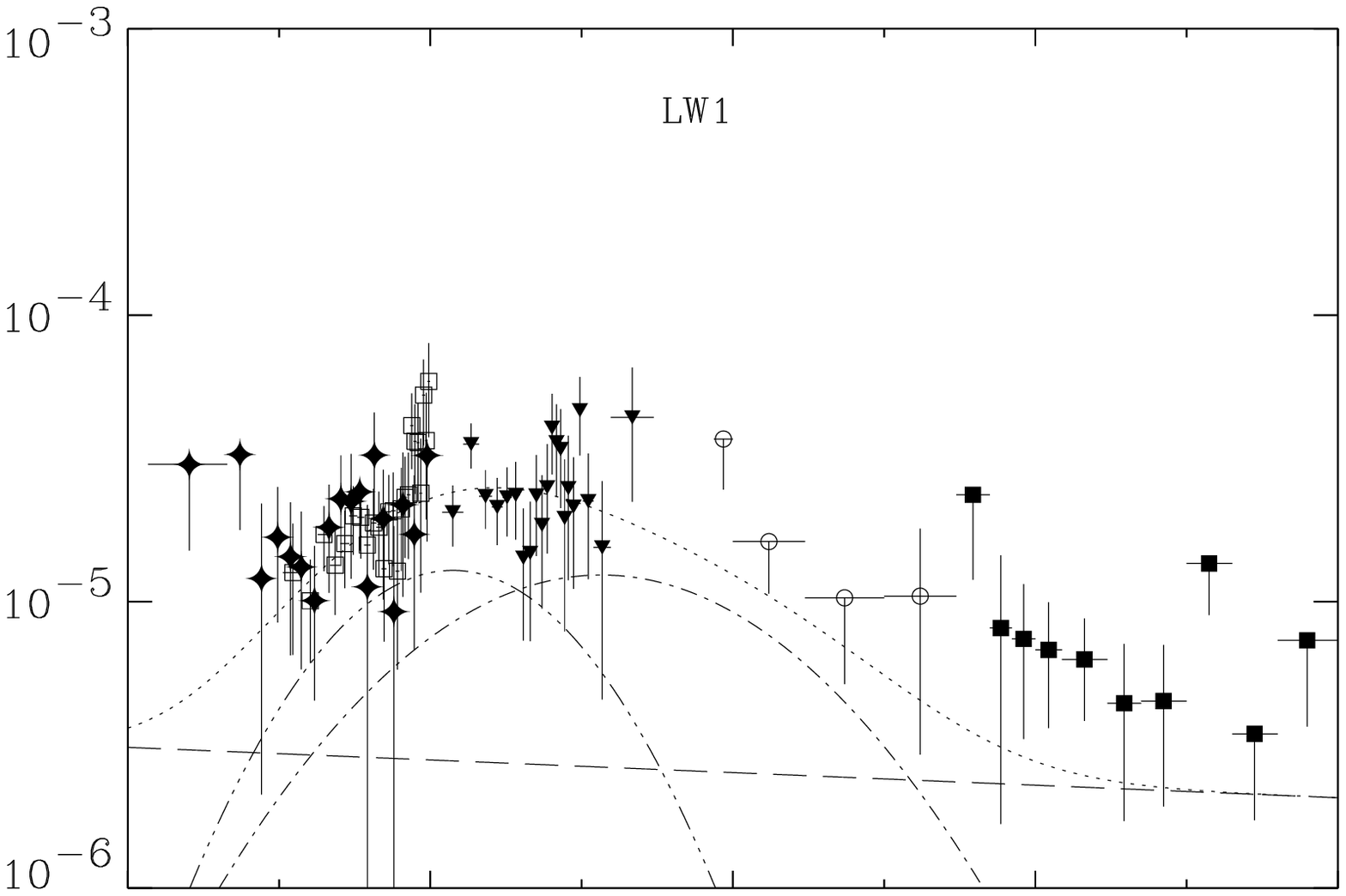,width=6.6cm,height=5.9cm}} \hspace{-0.9cm}
                               {\psfig{file=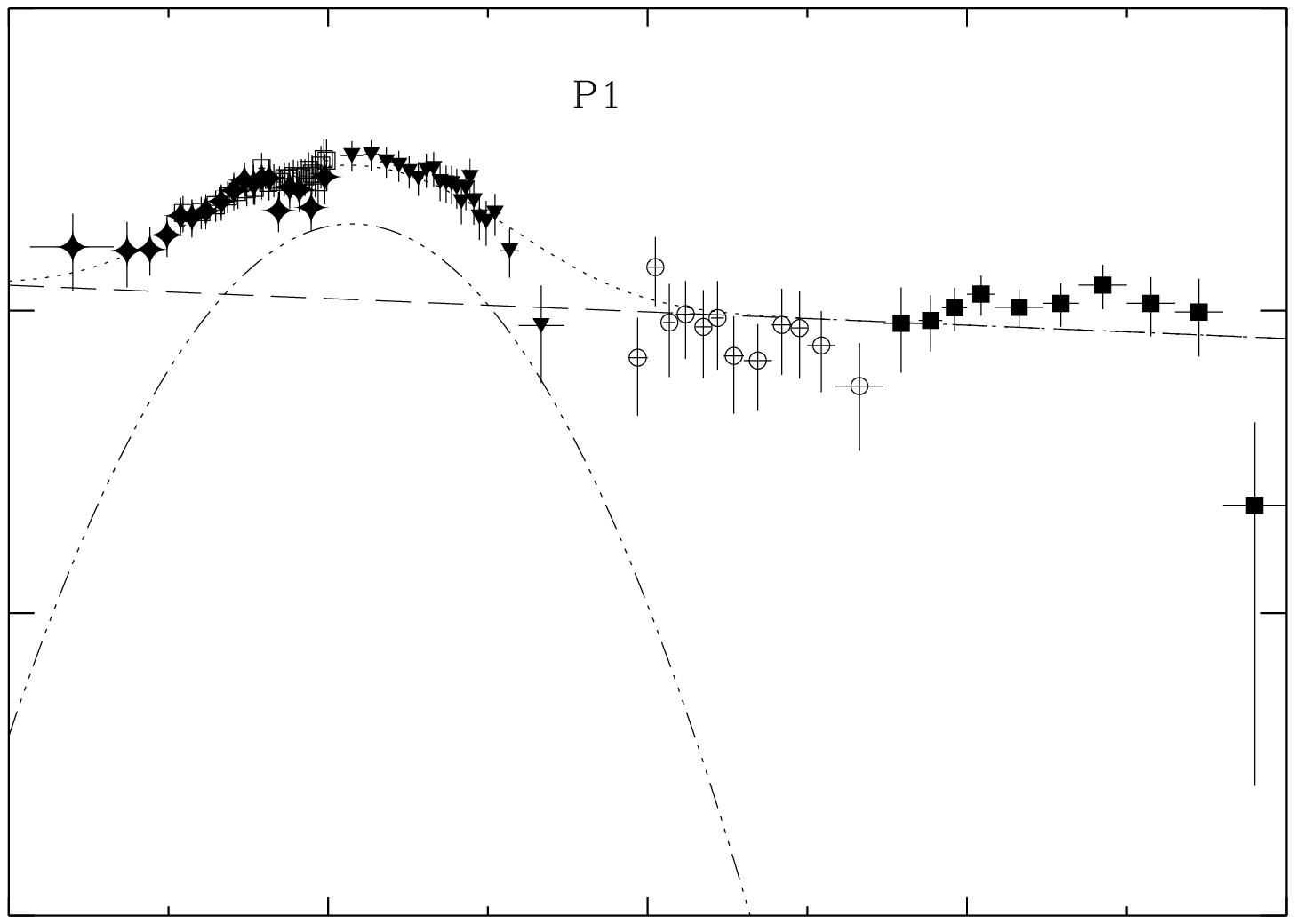,width=6.6cm,height=5.9cm}} \hspace{-0.9cm}
                               {\psfig{file=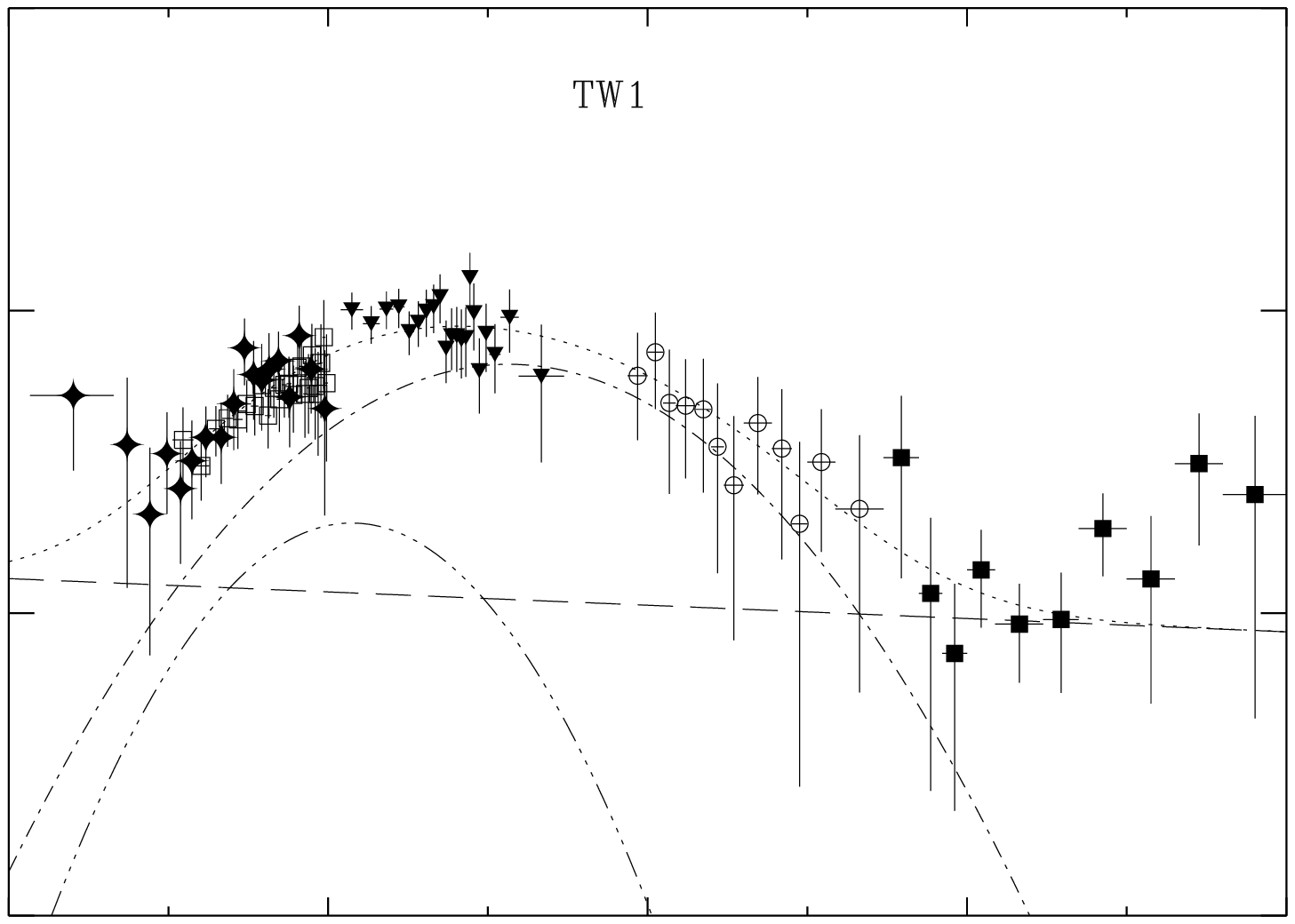,width=6.6cm,height=5.9cm}}  }
        \vspace{-0.7cm}

        \hbox{\hspace{-0.5cm}  {\psfig{file=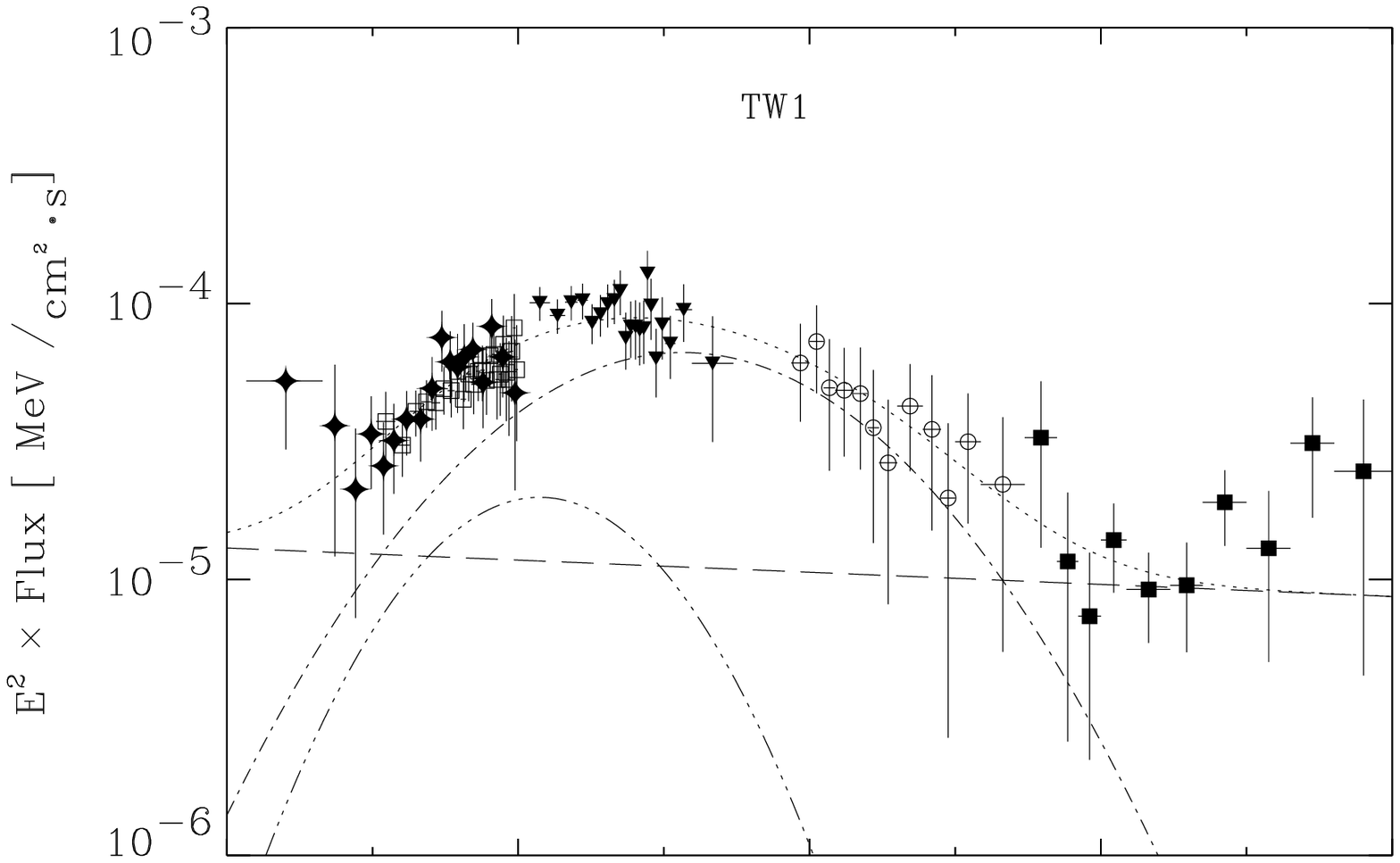,width=6.6cm,height=5.9cm}} \hspace{-0.9cm}
                               {\psfig{file=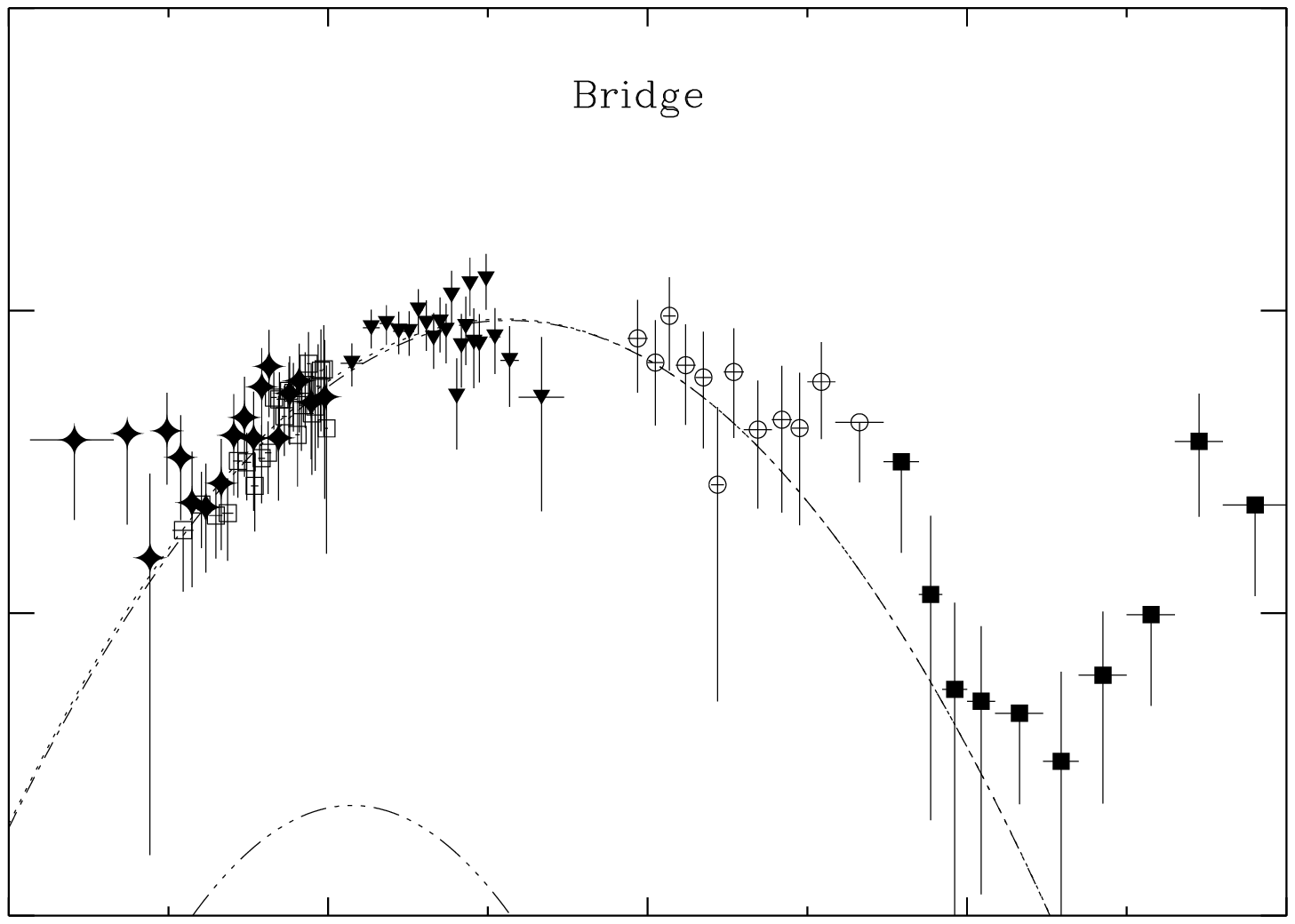,width=6.6cm,height=5.9cm}} \hspace{-0.9cm}
                               {\psfig{file=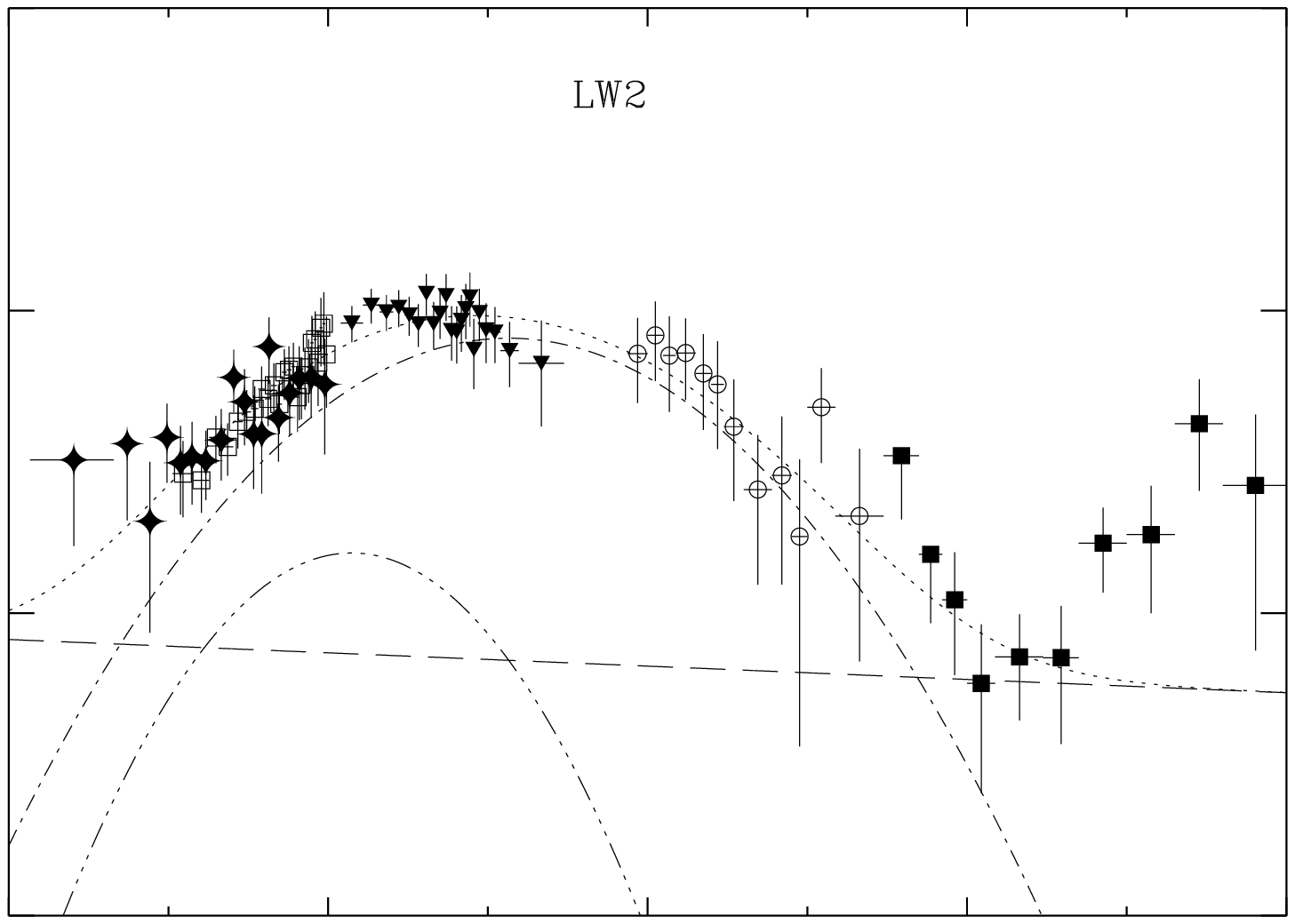,width=6.6cm,height=5.9cm}}  }

        \vspace{-0.7cm}

        \hbox{\hspace{-0.5cm}  {\psfig{file=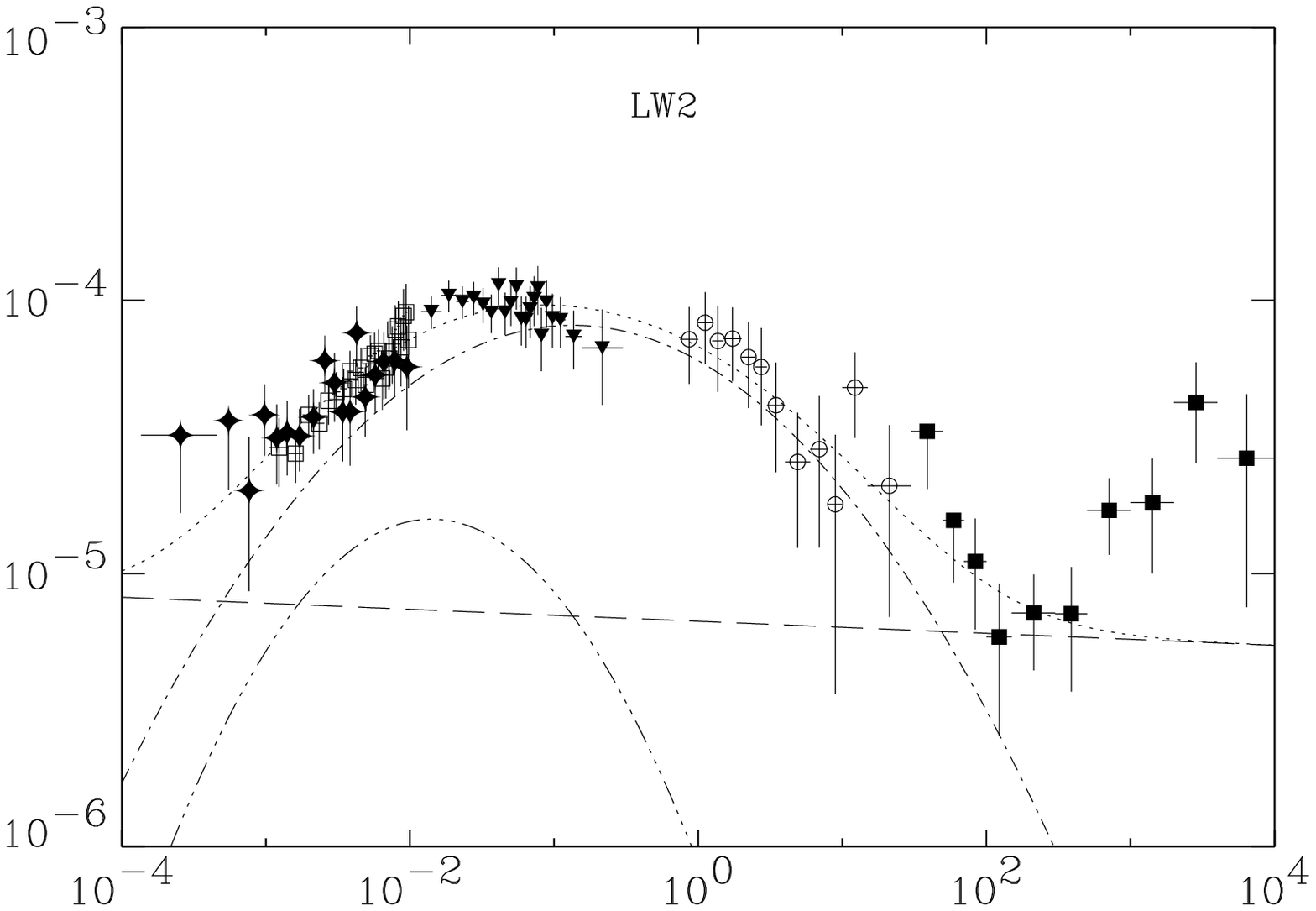,width=6.6cm,height=5.9cm}} \hspace{-0.9cm}
                               {\psfig{file=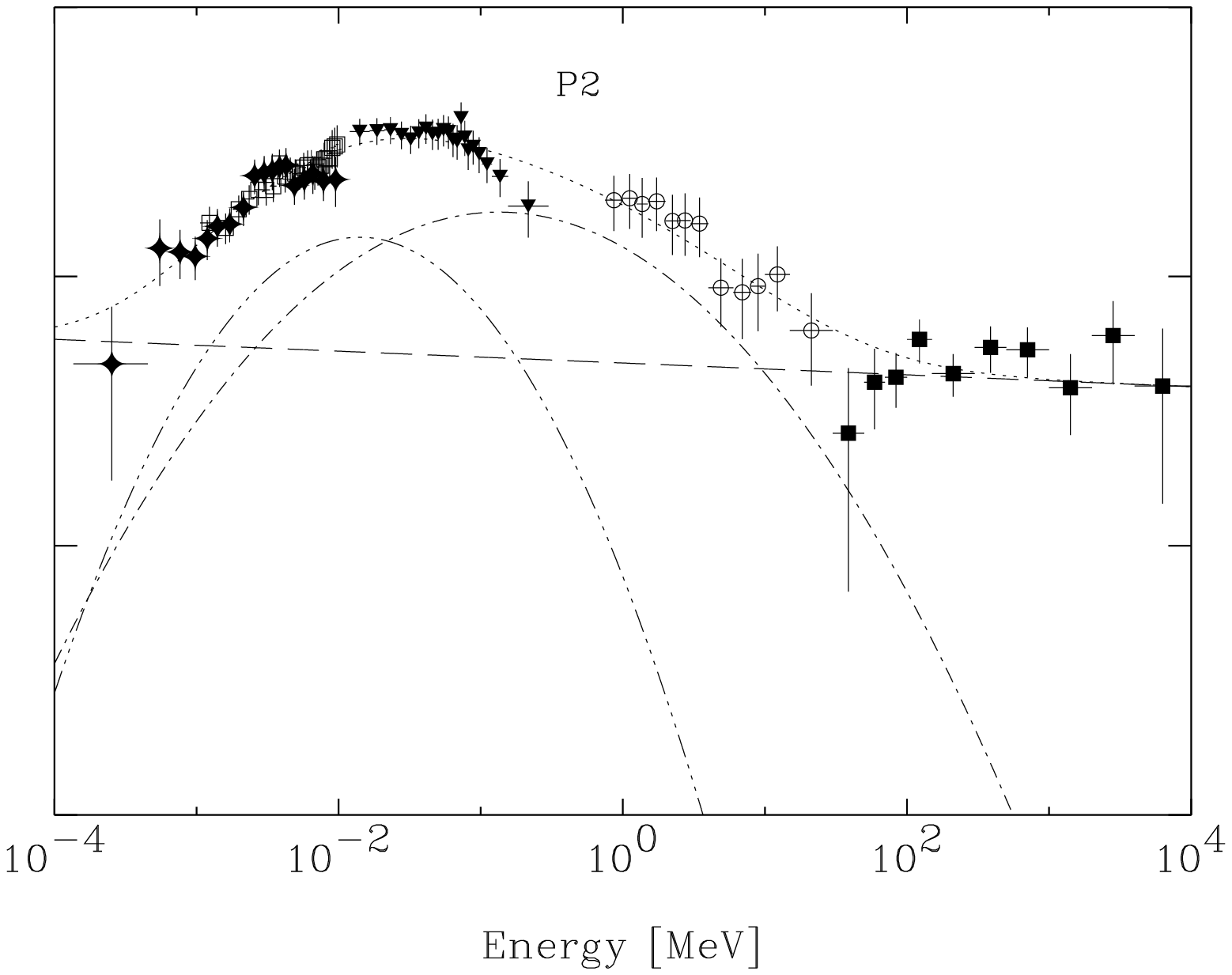,width=6.6cm,height=5.9cm}} \hspace{-0.9cm}
                               {\psfig{file=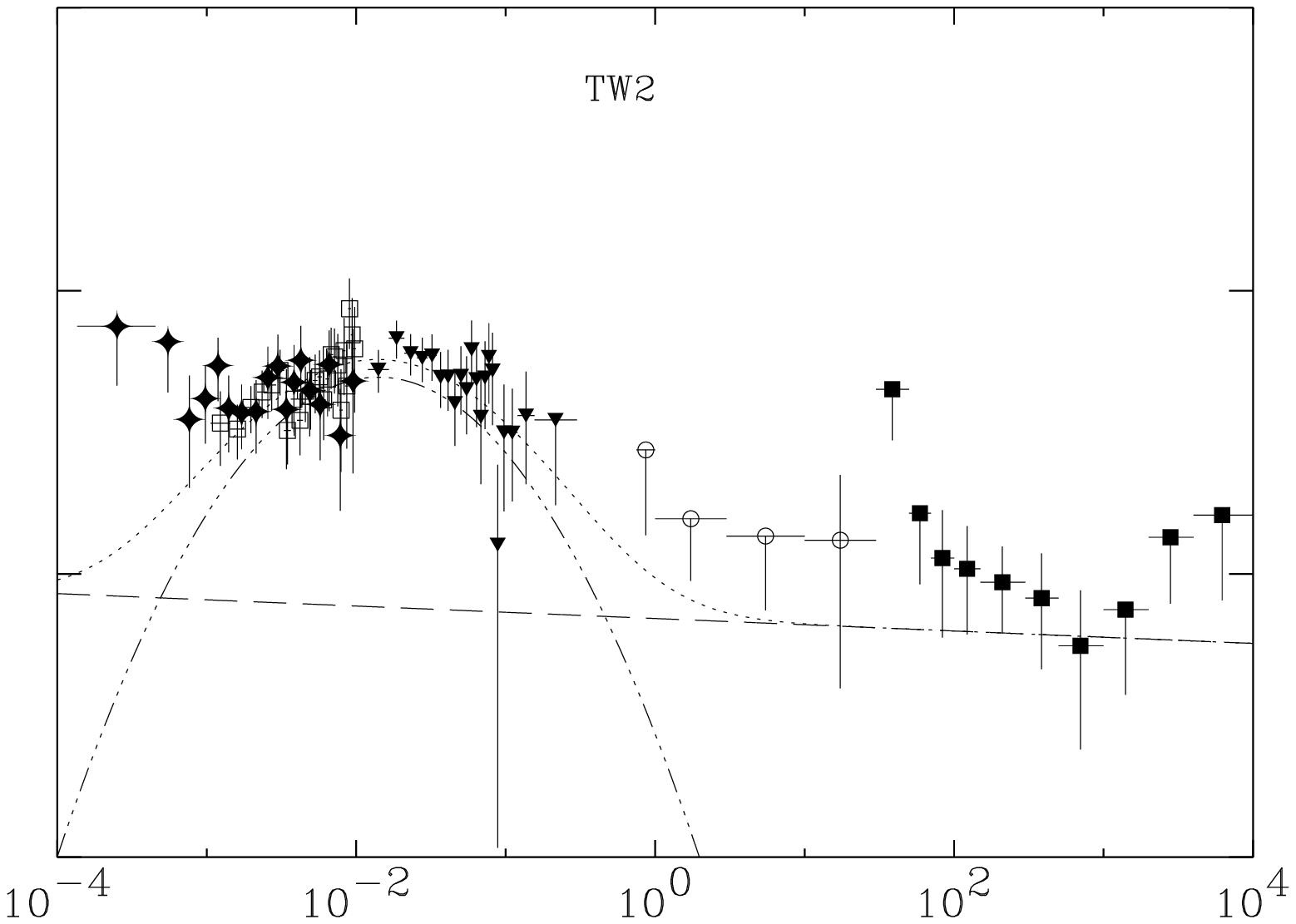,width=6.6cm,height=5.9cm}}  }
        \vspace{0.35cm}

                }
        \caption[]{The high-energy emission of the Crab pulsar in the 7 narrow pulse-phase intervals 
                   (see Table \ref{tab_pulsecomp}) from 0.1 keV up to 10 GeV. Two spectra (for the TW1 and LW2 phase intervals) 
                   are displayed twice, to facilitate a better visual comparison of the different 
                   spectra (see discussion). See Fig. \ref{fig:tp_spectrum} for the meaning of the symbols of the data points. 
                   The different contributions (P1 modified power-law model, dash-dot-dot-dot line; Bridge modified power-law 
                   model, dash-dot line; power-law model, dash line) to the composite model fits (dotted lines) have 
                   all been superimposed for the best-fit scaling parameters (histogram) shown in Fig. \ref{fig:scale_par}.
                   \label{fig:narrow_cmp_spectra}}
\end{figure*}}

\subsection{Spectral behaviour in the narrow pulse-phase intervals}
\label{section_narspc}

A similar broad-band spectral analysis (0.1 keV - 10 GeV), as presented in Sect. \ref{section_tpspec} for the Total Pulse interval,
has been performed for the 7 narrow pulse-phase intervals defined within the Total Pulse phase interval (see Table \ref{tab_pulsecomp}). 
We analyzed data from the BeppoSAX LECS, MECS and PDS and CGRO COMPTEL and EGRET instruments applying identical phase window selections. 
Since we planned to make empirical fits to the multi-instrument spectra, we wished to avoid being too sensitive to the systematic discrete jumps in the overlapping spectra of the BeppoSAX LECS, MECS and PDS and GRIS, as shown in Fig. \ref{fig:op_zoomin}.
In the spectra and analysis presented here, we added a representative $10\%$ systematic flux uncertainty to the statistical uncertainty 
in each flux measurement involved (for COMPTEL $15\%$ was used). We also repeated the total analysis normalizing the BeppoSAX data on the much
lower GRIS value, as well as on the average normalization value of the BeppoSAX instruments. Our results are not sensitive to
these different normalizations.

{\begin{figure*}[t]
   \hspace{0.3cm}
   \hbox{\hspace{-0.25cm}  {\psfig{file=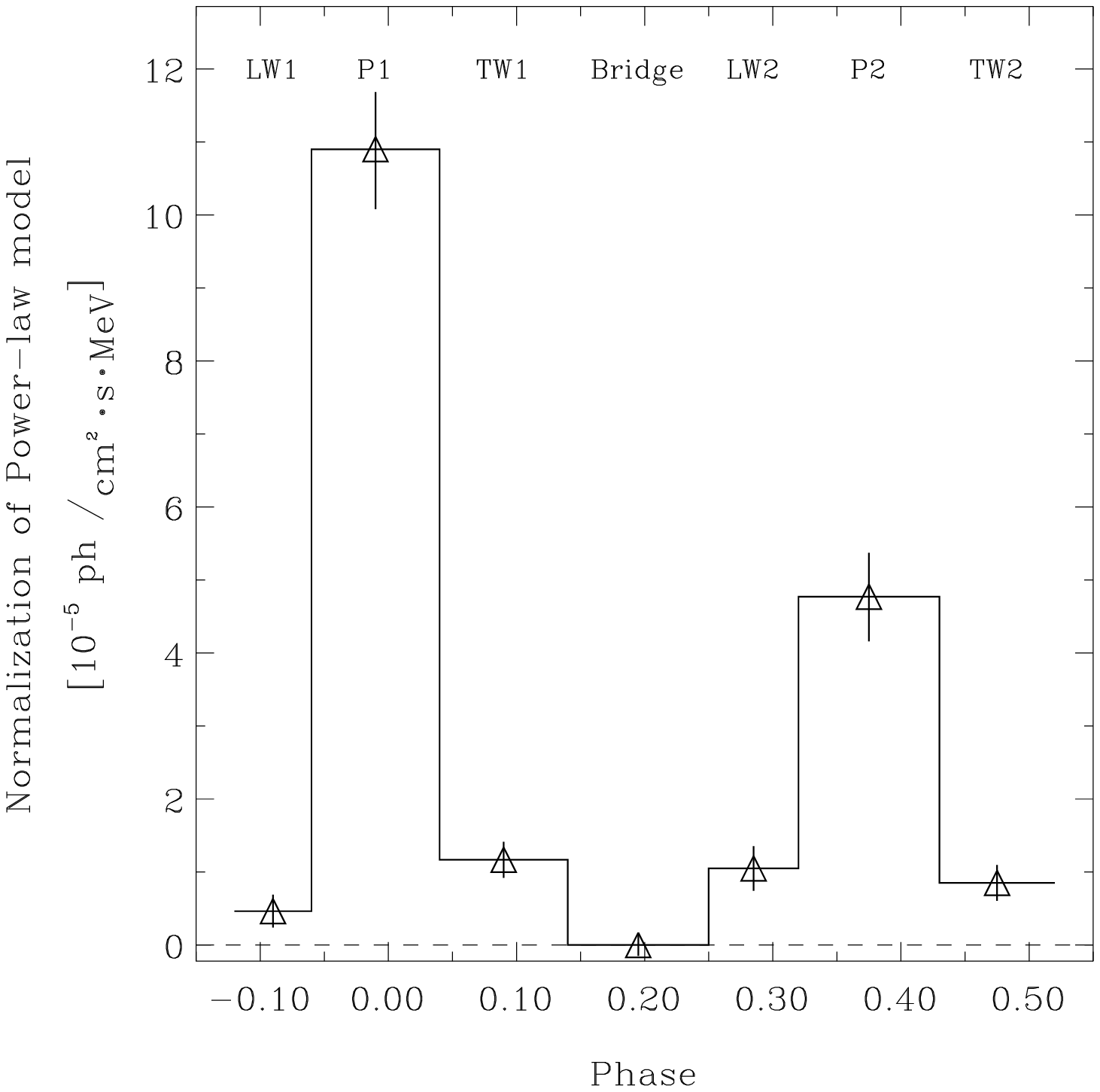,width=5.9cm,height=5.9cm}} \hspace{0.1cm} {\psfig{file=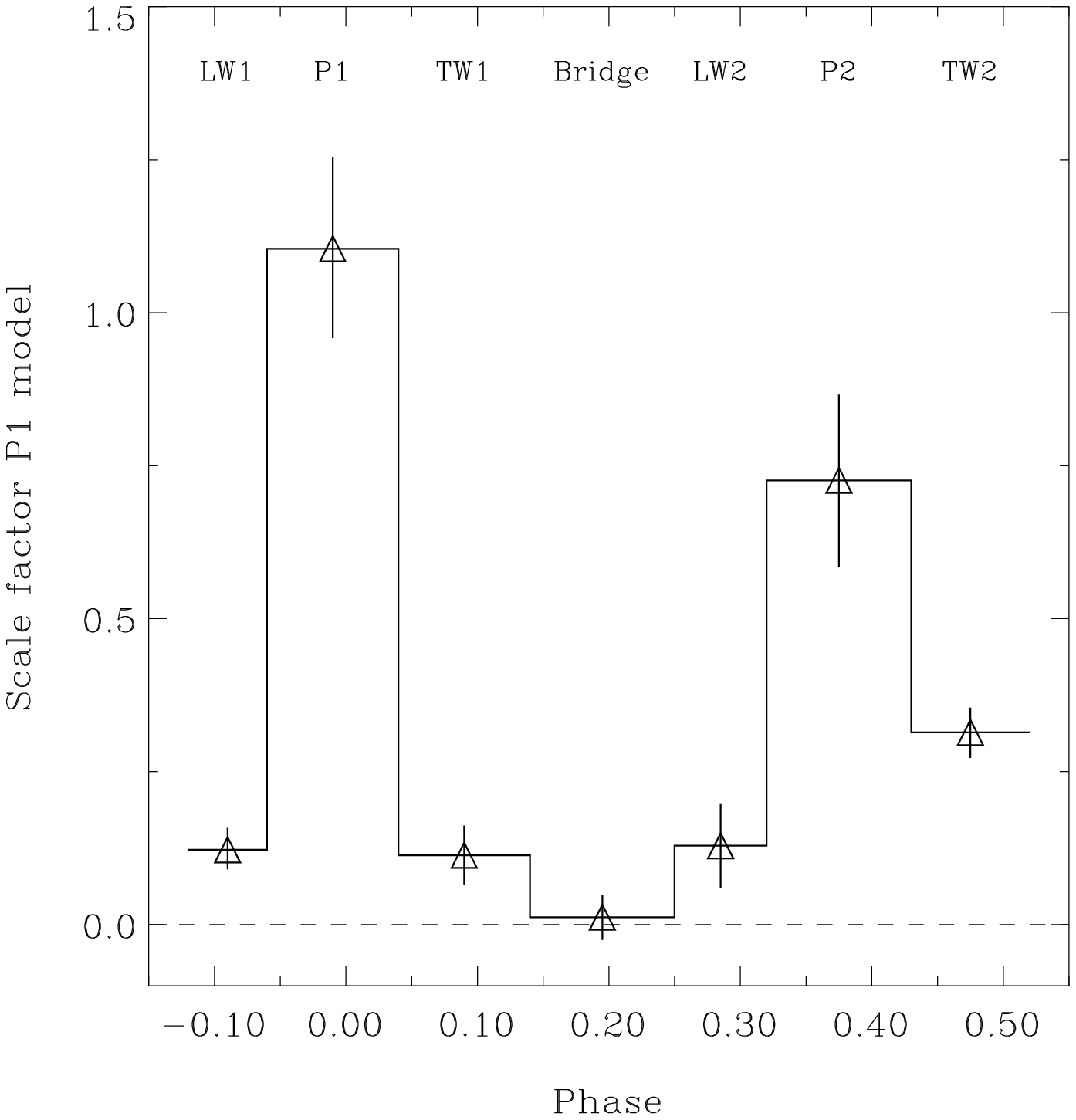,width=5.9cm,height=5.9cm}}
   \hspace{0.1cm} {\psfig{file=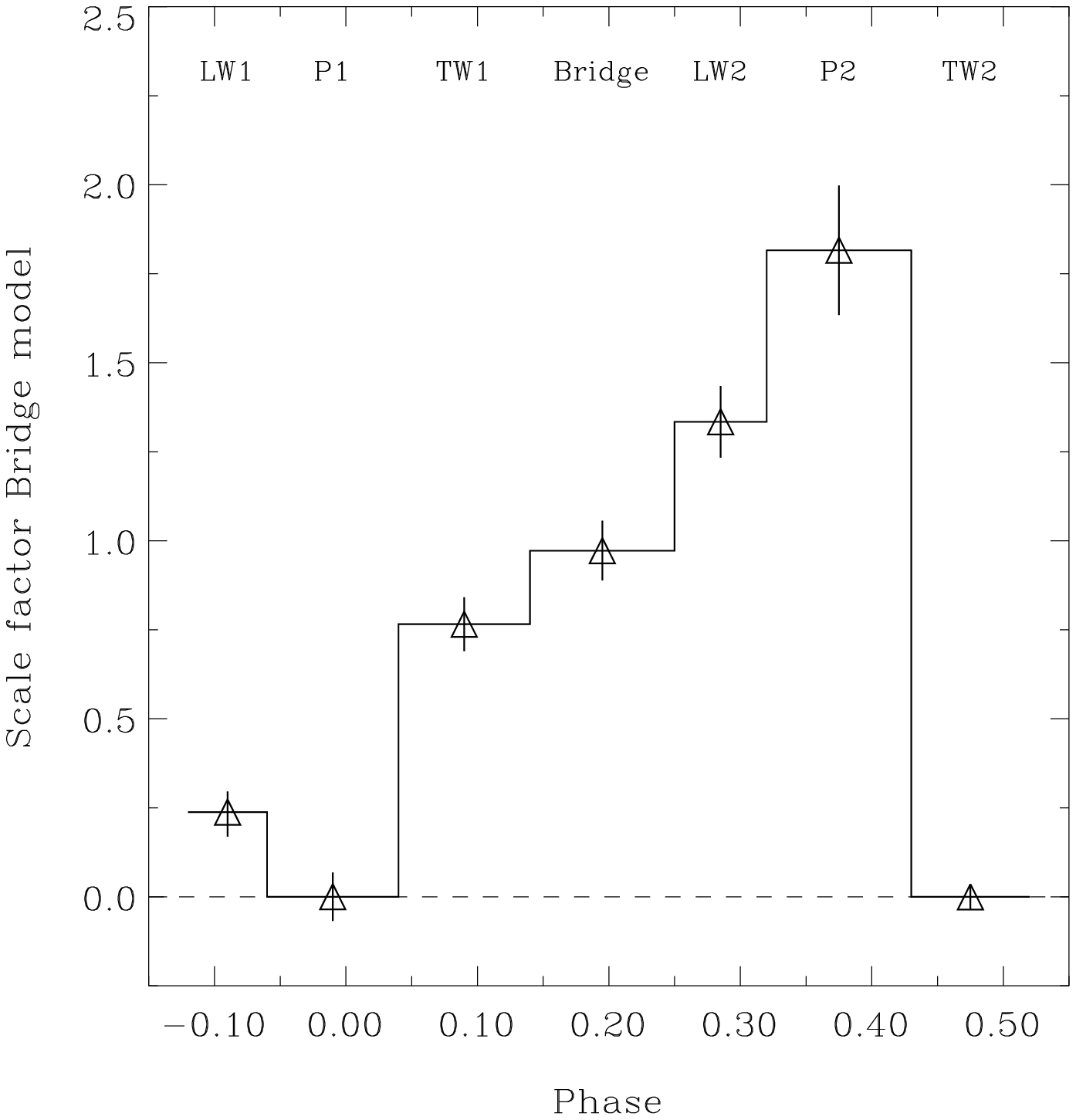,width=5.9cm,height=5.9cm}}
        }

   \vspace{0.2cm}
        \caption[]{Scale factors of the 3 empirical spectral models as a function of phase: left panel, power-law model scale factor;
                   middle panel, ``narrow" bump (P1 modified power-law component) scale factor; right panel, ``broad" bump (Bridge
                   modified power-law component) scale factor. \label{fig:scale_par}}
\end{figure*}}

The spectral results for the seven different intervals are shown in Fig. \ref{fig:narrow_cmp_spectra}\footnote{The phase-resolved
spectral data can be retrieved from {\it http://ws13.sron.nl:8080/personal/kuiper/data}}.
This compilation clearly shows that the spectral shape varies strongly with pulsar phase. For example, 
the spectra of the P1 and the Bridge intervals differ dramatically. Emission from the latter is hardly discernible at energies below 
1 keV and above 100 MeV; this emission is confined roughly between these energies in a broad ``bump'' shaped fashion in this $E^2\times F$ representation. 
On the other hand, the emission in the P1 interval remains very strong in the $\gamma$-ray domain above 1 MeV (COMPTEL and EGRET 
data), exhibiting a power-law photon distribution up to a spectral break at GeV energies. Extrapolation of such
a power-law spectral shape to X-ray energies, reveals a narrower ``bump'' shaped excess above the power-law extrapolation, 
with a maximum power output well below 50 keV. 
The P2 spectrum is rather similar to the spectrum of P1, but a spectral component similar in shape to that of the Bridge 
interval seems to enhance the P2 spectrum at MeV energies, relative to the P1 spectrum. Finally, the Trailing Wing 1 (TW1) 
and Leading Wing 2 (LW2) spectra are amazingly similar in shape, and appear to be some mixture of shapes of P1 and the 
Bridge intervals. This should not be a surprise, because the adopted separation in phase intervals (Table \ref{tab_pulsecomp}) 
will most likely not coincide exactly with genuine physical components (different dominating production mechanisms and/or 
production sites in the Crab magnetosphere). However, the vastly different spectral behaviour exhibited in the Bridge phase interval suggests a physically distinct emission component as proposed earlier by e.g. Knight (1982) and \cite{grcrab_hasinger_two}.
 

\subsection{Parametrization of the emission in the narrow pulse-phase intervals}
\label{sect:spc_bhv}
 
Exploiting our high statistics and eight-decades wide high-energy phase-resolved spectra, we made an attempt to 
empirically disentangle in phase and energy space underlying physical components. Assuming again that the 
Bridge spectrum represents the shape of a distinct emission component, we modelled its spectral behaviour in terms 
of two spectral components, a ``modified'' power-law (mpl) with an energy dependent index and a simple power-law (pl): 
$F= F^{\hbox{\rm \scriptsize mpl}}+F^{\hbox{\rm \scriptsize pl}}= \alpha\cdot E^{-(\beta+\gamma\cdot \ln E)} + 
\alpha_{\hbox{\rm \scriptsize pl}}\cdot E^{-\beta_{\hbox{\rm \scriptsize pl}}}$. 
In this formula $F$ denotes the photon flux in units ph/cm$^2$s MeV, while $E$ is given in MeV. As expected, the
normalization parameter $\alpha_{\hbox{\rm \scriptsize pl}}$ for the simple power-law component was 
consistent with zero, although a weak level of high-energy $\gamma$-ray emission is measured up to GeV energies. 
The same approach has been followed for the emission in the P1 interval. The resulting simple 
power-law component, describing entirely the high-energy end of the spectrum, has an index $2.022\pm 0.014$. We note that this 
power-law spectrum has to break near the boundaries of the energy window shown in Fig. \ref{fig:narrow_cmp_spectra} 
(see also Fig. \ref{fig:tp_spectrum}).
The resulting best fit values for the ``modified'' power-law components of the P1 and Bridge emissions are given 
in Table \ref{table:cmp_fits}. From these values we can derive the following positions for the maxima (in the $E^2\times F$ flux 
representation): $14.0\pm 1.1$ keV and $135\pm 15$ keV for the P1 and Bridge phase intervals, respectively.
The widths of the ``modified'' power-law components are specified by the FWHM values in the $^{10}\log(E)$ domain and 
are about $1.81$ and $2.65$ for the P1 and Bridge intervals, respectively. With this approach we have identified three 
distinctly different spectral shapes, which can describe the P1 and Bridge spectra. 

\begin{table}[h]
\caption[]{\label{table:cmp_fits} Best fit parameter values for the ``modified'' power-law \\
                                  and power-law components of the P1 phase interval and for the\\
                                  ``modified'' power-law of Bridge phase interval}
\begin{flushleft}
\tabcolsep=0.75\tabcolsep
\begin{tabular}{lccc}
\hline\noalign{\smallskip}
Component  & $\alpha$      & $\beta$  & $\gamma$      \\
           & $(\hbox{\rm ph}/\hbox{\rm cm}^2 \hbox{\rm s} \hbox{\rm MeV} )$      &   &       \\
\noalign{\smallskip}
\hline\noalign{\smallskip}
P1-mpl     & $(1.06\pm 0.07)$E-5    &  $3.361\pm 0.014$     & $0.159\pm 0.003$   \\
P1-pl      & $(9.98\pm 0.69)$E-5    &  $2.022\pm 0.014$     &                    \\
\noalign{\smallskip}
Bridge-mpl & $(7.09\pm 0.51)$E-5    &  $2.298\pm 0.017$     & $0.074\pm 0.003$   \\
\noalign{\smallskip}
\hline\noalign{\smallskip}
\end{tabular}
\tabcolsep=1.3333\tabcolsep
\end{flushleft}
\end{table}

In the next step we made an attempt to describe the measured spectral distributions in the narrow
pulse-phase intervals (npi) in terms of just these 3 models each with a free scaling parameter:
$F^{\hbox{\rm \scriptsize npi}}= a \cdot F^{\hbox{\rm \scriptsize Bridge-mpl}} + b \cdot F^{\hbox{\rm \scriptsize P1-mpl}}
   + c \cdot E^{-\beta_{\hbox{\rm \scriptsize P1-pl}}}$

Interestingly, the resulting fits are very satisfactory for all phase intervals as shown in Fig. \ref{fig:narrow_cmp_spectra} 
in which the composite model (dotted lines) and the individual components are superimposed on the measured spectra. 
The fit characteristics for each phase interval are shown in Table \ref{table:mdl_fits}. The $\chi_{\nu}^2$ values of the 
fits indicate acceptable spectral descriptions in all cases. The inclusion of $10$-$15\%$ systematic uncertainties in the 
flux measurements, however, makes a straightforward assessment/interpretation of the $\chi_{\nu}^2$ values difficult.

\begin{table*}[t]
  \caption[]{\label{table:mdl_fits} Best fit scale factors $(a,b,c)$ for the narrow phase intervals\\}

  \begin{flushleft}
    \begin{tabular}{lcccc}
      \hline\noalign{\smallskip}
      Interval  & $a$      & $b$  &   $c$  &  $\chi_{\nu}^2 \ (\chi^2/n_{\hbox{\rm \scriptsize dof}})$   \\
      \noalign{\smallskip}
      \hline\noalign{\smallskip}
      \noalign{\smallskip}
LW1 & $0.130^{+0.024\ +0.008}_{-0.029\ -0.008}$  & $0.067^{+0.011\ +0.008}_{-0.009\ -0.008}$ 
    & $(\ \ 2.53^{+1.01\ +0.22}_{-1.00\ -0.22})$E-6  & $0.94\ \left( 69.83 / (77 - 3) \right) $\\
      \noalign{\smallskip}
P1  & $0.000^{+0.061\ +0.001}_{-0.061\ -0.001}$  & $1.004^{+0.050\ +0.086}_{-0.047\ -0.086}$ 
    & $(99.09^{+4.59\ +2.52}_{-4.93\ -2.52})$E-6 & $0.66\ \left( 53.80 / (85 - 3) \right) $\\
      \noalign{\smallskip}
TW1 & $0.696^{+0.052\ +0.017}_{-0.052\ -0.017}$  & $0.103^{+0.019\ +0.025}_{-0.019\ -0.025}$ 
    & $(10.62^{+1.85\ +0.39}_{-1.86\ -0.39})$E-6  & $0.46\ \left( 38.07 / (85 - 3) \right) $\\
      \noalign{\smallskip}
Bridge & $0.972^{+0.051\ +0.033}_{-0.050\ -0.033}$  & $0.012^{+0.018\ +0.019}_{-0.018\ -0.019}$ 
    & $(\ \ 0.01^{+1.40\ +0.07}_{-1.41\ -0.07})$E-6 & $0.76\ \left( 62.48 / (85 - 3) \right) $\\
      \noalign{\smallskip}
LW2 & $0.849^{+0.048\ +0.016}_{-0.048\ -0.016}$  & $0.082^{+0.017\ +0.027}_{-0.017\ -0.027}$ 
    & $(\ \ 6.68^{+1.59\ +0.35}_{-1.60\ -0.35})$E-6 & $0.59\ \left( 48.22 / (85 - 3) \right) $\\
      \noalign{\smallskip}
P2 & $1.816^{+0.118\ +0.064}_{-0.122\ -0.064}$  & $0.726^{+0.047\ +0.093}_{-0.048\ -0.093}$ 
    & $(47.71^{+4.05\ +1.95}_{-4.15\ -1.95})$E-6 & $0.48\ \left( 39.61 / (85 - 3) \right) $\\
      \noalign{\smallskip}
TW2 & $0.000^{+0.027\ +0.000}_{-0.027\ -0.000}$  & $0.257^{+0.016\ +0.017}_{-0.017\ -0.017}$ 
    & $(\ \ 6.96^{+1.53\ +0.48}_{-1.53\ -0.48})$E-6 & $0.83\ \left( 61.10 / (77 - 3) \right) $\\
      \noalign{\smallskip}

      \noalign{\smallskip}
      \hline\noalign{\smallskip}
    \end{tabular}
  \end{flushleft}

\end{table*}

In Table \ref{table:mdl_fits} two types of error estimates are presented for each scaling parameter. The first type
is associated with the (asymmetric) statistical uncertainty in the scale parameter using the best fit estimates for the 
parameters describing the shape of the 3 models  i.e. $\beta_{\hbox{\rm \scriptsize P1-mpl}},\gamma_{\hbox{\rm \scriptsize P1-mpl}}, 
\beta_{\hbox{\rm \scriptsize Bridge-mpl}},\gamma_{\hbox{\rm \scriptsize Bridge-mpl}}$ and $\beta_{\hbox{\rm \scriptsize P1-pl}}$ 
(see Table \ref{table:cmp_fits}). The second type is related to the systematic uncertainty in the fitted scale parameter and has been determined by varying the shape parameters of the 3 models within their $\pm 1\sigma$ errors.
The range over which the scale parameters vary is indicative for the systematic uncertainty due to uncertainties in the shape
of the 3 model fit functions. 

The fit results for the 3 scale parameters are visualized in Fig. \ref{fig:scale_par}. These scale parameters have 
been normalized to the emission in the Bridge interval because of the different phase extents of the intervals. 
Fig. \ref{fig:scale_par} shows effectively the ``light curves'' of these model components: two components are clearly 
related to the emission in the two main pulses (the power-law component and the ``narrow bump'') and the
``Bridge component'' or ``broad bump'' extends apparently from the LW1 till under P2 in a triangular shape. 

The ``light curves" in Fig. \ref{fig:scale_par} can also be clearly discerned in the 9 panels of Fig. \ref{fig:narrow_cmp_spectra}.
The upper row centered on P1 shows how the components of P1 extend into the wings; the middle row shows how the Bridge spectrum
dominates in all three intervals and how the power-law component and the ``narrow bump" contributions are symmetrically distributed
on either side of the Bridge interval. Finally, the lowest row centered on P2 looks very much like the upper row, but the Bridge
spectral component reaches a maximum value in P2. We have apparently succeeded in identifying likely genuine underlying 
physical components in phase and energy space.
 

\subsection{Enhanced high-energy $\gamma$-ray emission in the LW2 interval}

The most apparent and significant ($4.3\sigma$ for energies above 300 MeV) deviation from the composite fits in Fig. 
\ref{fig:narrow_cmp_spectra} is visible in the LW2 spectrum in the EGRET range above 100 MeV. \cite{grcrab_fierro_two} reported 
for this phase interval the hardest pulsed $\gamma$-ray spectrum (photon index $1.69\pm 0.08$) fitting CGRO Cycles 0-III
EGRET data.
For our (timing) analysis we used CGRO Cycle 0-VI EGRET data, almost doubling the statistics for energies above 1 GeV\footnote{
for energies lower than 1 GeV the increase in statistics is less due to the energy and time dependent EGRET sensitivity
caused by spark-chamber gas aging.}

To verify whether the phase distributions in the GeV energy range also show evidence for a separate hard {\em spectral} component 
in front of P2 we produced the pulse profiles of Fig. \ref{fig:he_prof}: the 1-10 GeV pulse profile superimposed on the 100-300 
MeV profile for EGRET Cycle 0-VI observations. The two distributions are normalized on the P1 phase interval.
A clear increase of emission in the LW2 phase interval (0.25-0.32) is visible for the highest energies, but the enhancement
seems to extend to the maximum in the P2 phase interval, however constituting a minor fraction in the latter. The effect can be 
interpreted as a phase shift of the second pulse with increasing energy, but also as a ``new" spectral component (phases $\sim$ 0.2-0.4) 
with respect to the composite fits with three spectral components as shown in Fig. \ref{fig:narrow_cmp_spectra}. 
\cite{grcrab_murthy} and \cite{grcrab_eiken} discussed variations in the intra-peak phase separations as a function 
of energy. A further phase shift above 1 GeV is consistent with the reported trend. 

Considering the enhancement a new spectral component then its spectrum must be even harder than derived by \cite{grcrab_fierro_two}. 
Fitting for the LW2 phase interval another power-law model with both a free normalization and a free 
index on top of the model composed of the broad and narrow spectral components and the power-law component with a fixed index of 
$2.022$, we find for the additional power-law model a very hard photon index of $1.44^{+0.05}_{-0.04}$.
However, our data are not of sufficient quality to discriminate between an additional very hard power-law component, which 
must break somewhere above 10 GeV to be consistent with the non-detections of pulsed emission at TeV energies, or for example an 
additional high-energy ``bump". This can be studied in detail with the next generation high-energy $\gamma$-ray telescopes 
AGILE and particularly GLAST (see e.g. {\it http://glast.gsfc.nasa.gov}), which is $\sim 30$ times more sensitive than EGRET 
over a much wider energy interval extending to 300 GeV. 

{\begin{figure}[t]
  \hspace{0.3cm}
  {\psfig{file=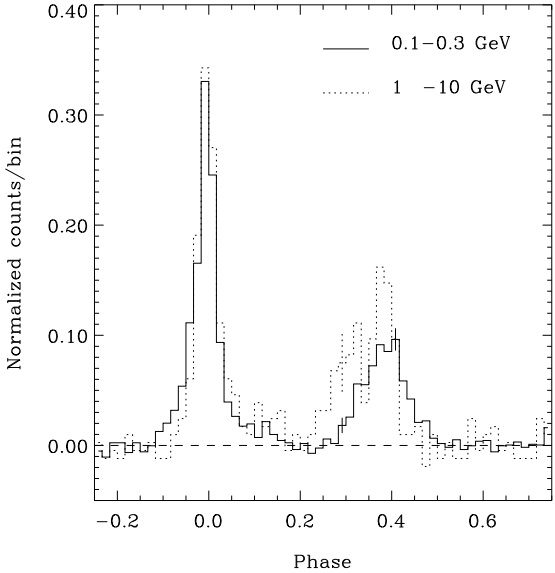,width=8cm,height=8cm}}
  \caption[]{EGRET pulse profiles (60 bins) using CGRO EGRET Cycle 0-VI data: 1-10 GeV, dotted line;  
             100-300 MeV, solid line. The profiles are normalized on their emission 
             in P1. Typical error bars are indicated for both profiles. A clear increase is visible in the 
             LW2 phase interval (0.25-0.32) for the 1-10 GeV energy interval with respect to the 100-300 MeV interval. 
             \label{fig:he_prof}}
\end{figure}}


\section{Summary and discussion}

In this work we derived final COMPTEL pulse profiles and spectra for the Crab pulsar and nebula at medium 
$\gamma$-ray energies (0.75-30 MeV) using data collected over the 9 year mission of NASA's Compton Gamma-Ray
Observatory. 

Due to the high counting statistics over the total 0.75-30 MeV interval, we were able to show a clear morphology 
change of the pulse profile as a function of energy, providing clear evidence for drastic spectral variations with 
pulsar phase over the COMPTEL energy window. Indications for such variations were found in the earlier 
COMPTEL analysis by \cite{grcrab_muchone}.

Using our large COMPTEL data base we derived an improved Crab nebula spectrum, which has
a power-law spectral shape between 0.75 and 30 MeV with index $2.227 \pm 0.013$.
Also our new COMPTEL spectrum for the total pulsed emission (nebula/DC emission
subtracted) can be described with a power-law spectral shape between 0.75 and
30 MeV with index $2.24 \pm 0.04$. If the indication for enhanced emission
in the 10-15 MeV interval is genuine, then the index becomes, $2.35 \pm 0.06$.

These improved COMPTEL findings have been put in a much broader context by 
including in our analysis data from instruments sensitive at the neighbouring X-ray/soft 
$\gamma$-ray energies, particularly from the BeppoSAX LECS, MECS and PDS 
instruments, and at high $\gamma$-ray energies from CGRO EGRET. 

We compiled a new spectrum of the Crab pulsed emission from optical wavelengths up to the high-energy $\gamma$-rays at 10 GeV 
(Fig. \ref{fig:tp_spectrum}). This emission reaches a level of maximum luminosity per
decade in energy from $\sim$ 5 keV to 50 keV. Beyond this maximum a gradual softening sets in 
reaching a plateau (photon power-law index of $\sim 2$) near $\sim 30$ MeV which 
continues to $\sim 10$ GeV. Above $\sim 10$ GeV the spectrum 
must break rapidly in order to be consistent with the stringent TeV upper limits for pulsed emission.

Phase resolved spectral analysis can provide important constraints for pulsar 
modelling, particularly to help identifying different production mechanisms and sites
in the pulsars magnetosphere. Therefore, we derived consistently over a broad energy range
from 0.1 keV up to 10 GeV (BeppoSAX LECS, MECS, PDS, CGRO COMPTEL and EGRET) for seven narrow phase intervals
phase-resolved spectra. These spectra exhibited very different spectral shapes, most notably the spectra
for the narrow Bridge and Peak 1 intervals.
We could disentangle the pulsed emission in energy and phase space, exploiting  the vastly different 
spectral shapes, particularly over the COMPTEL energy window, by making empirical fits, and found 
that the pulsed emission can be described with 3 distinctly different spectral components:
\begin{description}
  \item{-1- }{a power-law emission component from $\sim$ 1 keV to $\sim$ 5 GeV, photon
  index $2.022 \pm 0.014$, which is present in the phase intervals of the two pulses.}
  \item{-2- }{a curved spectral component required to describe soft ($\leq 100$ keV) excess emission present in the 
  same pulse-phase intervals.}
  \item{-3- }{a broad curved spectral component reflecting the bridge emission from 0.1 keV to $\sim$ 10 MeV. 
  This broad spectral component extends in phase over the full pulse profile in an approximately triangular shape,
  peaking under the second pulse.}
\end{description}

Furthermore, in addition to the 3 spectral components the Leading Wing 2 (LW2)  phase interval exhibited a very hard
spectral component, most notably at GeV energies, which likely extends over the broader phase interval $\sim$ 0.2-0.4.

In a somewhat different approach and using only BeppoSAX data between 0.1 and 300 keV, Massaro et al. (2000)
identified recently two components: the first is the combination of the two components as described above under -1- and -2- 
and the second corresponds to the one described under -3-. In their narrower energy window
differences in spectral shapes can be well approximated by variations in power-law index.

Since the discovery of $\gamma$-ray emission from radio pulsars in the early seventies, 
the most popular and competing models attempting to explain the high-energy radiation from highly magnetized
rotating neutron stars can be divided in two distinct catagories: the so-called Polar Cap (PC) models 
and Outer Gap (OG) models. 

Detailed information on the PC models can be found in e.g.: Daugherty \& Harding (1982,1994,1996), Sturner \& Dermer (1994) 
and most recently \cite{grcrab_zhang}. PC models have problems in explaining the overall measured Crab pulsar characteristics. 
Most notibly, the large angles of $\sim 60\degr$ estimated for both the magnetic inclination $\alpha$ and the viewing angle $\zeta$, 
the angle between the spin axis and the observer's line of sight, from radio and optical/UV observations cannot be reconciled 
(see for recent publications e.g. Graham-Smith et al. 1996, Moffett \& Hankins 1999 and Everett \& Weisberg 2001).

OG models have no difficulties with the large $\alpha$ and $\zeta$ angles estimated for the Crab, as is clearly shown by 
\cite{grcrab_chiang}. In these models the acceleration of charged particles and production of high-energy radiation takes
place in charge depleted gaps between the null-charge surface, defined by \mbox{\boldmath{$\Omega \cdot B$}} $=0$ with 
\mbox{\boldmath{$B$}} the local magnetic field, and the light cylinder (with radius $R_{lc}=c/\Omega$) above the last
closed field lines. For early papers and later refinements see: Cheng et al. (1986a,b), \cite{grcrab_ho}, \cite{grcrab_chiang}, 
\cite{grcrab_romani}, \cite{grcrab_romanitwo} and Yadigaroglu (1997). 
Recently, \cite{grcrab_chengthree} presented a three-dimensional outer gap model building on the work of Romani and
co-workers. The emission patterns from these outer gap models resemble fan beams, and double peak profiles with
(strong) bridge emission can commonly be generated for the cases that emission is seen from only one pole, e.g. 
\cite{grcrab_romanitwo}, as well as from both poles, \cite{grcrab_chengthree}.

Early attempts to model the Crab Total Pulse spectrum in an OG scenario using various radiative processes like the curvature, 
synchrotron and inverse Compton radiation mechanisms, assumed to play a key role in the outer gap physics, were made by 
Cheng et al. (1986b) and \cite{grcrab_ho}. The latter employed a self-consistent iterative procedure with one varying 
parameter, the ratio of gap height and curvature radius of the field lines, and it is interesting to note that the calculated 
high-energy spectrum bears reasonably good overall simularity with the observed one (cf. Fig. 4 of \cite{grcrab_ho} with Fig. 
\ref{fig:tp_spectrum} of this paper). Ulmer et al. (1995) compared the outer gap model of Ho with an early CGRO spectrum of 
the Crab (combining OSSE, COMPTEL and EGRET spectra) and found also good overall agreement.
Chiang \& Romani (1994) made refinements to the above calculations and attempted to model the Crab Total Pulse spectrum and 
the spectral variation with phase, which they considered to be a clear mapping of location in the magnetosphere to pulse-phase. 
They divided the outer gap in different sub-zones taking into account the transport of radiation and particles 
from sub-zone to sub-zone. Convergence to a self-consistent solution, however, resulted in spectra significantly lacking photon 
flux below several GeV. Romani (1996) described a revised picture of gap closure and radiation physics in the outer 
magnetosphere to overcome difficulties in the schema of Cheng et al. (1986b), and also addressed spectral variations with 
pulsar phase from the optical to the high-energy $\gamma$-ray spectrum. 
For the Crab pulsar, he made some qualitative statements on the expected spectral properties. He expects a significant contribution
of synchrotron photons to the high-energy $\gamma$-ray flux. This could probably explain the observed underlying 
power-law component from soft X-rays up to high-energy $\gamma$-rays, although the observed photon index of $\sim 2$ is 
considerably softer than the expected value of $\sim 1$.

Building on the work of Romani and co-workers, \cite{grcrab_chengthree} also used a three-dimensional pulsar magnetosphere to 
study the geometry of outer magnetospheric gap accelerators. However, the physics of both models is strikingly different. For 
the single outer gap model (Chiang \& Romani 1994), the emission comes from the outward direction in an outer gap above one 
pole; the emission regions for the two peaks of the pulse profile are those close to the null-charge surface and to the light 
cylinder radius, respectively.
In the model of \cite{grcrab_chengthree}, photon emission consists of emission outward and inward from regions in outer 
gaps above both poles, the gaps being limited along the azimuthal direction by $e^{\pm}$ pair production of inward-flowing
photons from the outer gap. It is shown that both models can produce the same (Crab-like) pulse profiles. 
\cite{grcrab_chengthree} also calculated phase-resolved spectra of the Crab pulsar. They determined the locations of the 
emission regions in the outer gaps in the open field line zone of the Crab magnetosphere as a function of pulse-phase. It can 
be seen in their Fig. 9 that high-energy emission from the P1 interval is produced high in the magnetosphere ($0.8 < r/R_{lc} 
< 1.0$) where curvature radiation dominates, resulting in a spectrum which extends to the GeV regime. In the interval between 
the pulses (TW1 and Bridge interval) high-energy radiation is predominantly produced deep in the magnetosphere where a soft 
synchrotron component is expected to dominate, roughly in accordance with the observations. Moving towards the P2 interval, 
emissions from regions high and low in the magnetosphere contribute, resulting in a overall spectrum composed of a hard 
curvature component and a much softer synchrotron component. 
In the P2 interval high-energy radiation is coming, in essence, from emitting regions extending from $\sim 0.2 R_{lc}$ to $\sim 
1.0 R_{lc}$ which gives rise to a hard spectral component extending into the GeV domain and a soft component.
Crossing the bridge interval moving from P1 to P2 a gradual decrease from $\sim 0.6 R_{lc}$ to $\sim 0.2 R_{lc}$ is seen for 
the lower bound of the emission region in the outer gap. According to Eq. 33 of \cite{grcrab_chengthree} the dominating 
synchrotron emission from these regions deep in the outer gap in the pulsar's magnetosphere becomes increasingly intense 
moving towards P2, because the magnetic field strength becomes stronger deeper in the magnetosphere. This offers an 
explanation for the observed phase dependence of the bridge/broad bump spectral component, shown in Fig. \ref{fig:scale_par}. 
Beyond P2 (and before P1) this soft synchrotron component should be absent which is in agreement with the observations. Thus 
the model proposed by \cite{grcrab_chengthree} seems to provide a viable and promising theoretical description of the physics 
responsible for the production of the Crab high-energy radiation with characteristics as shown in Figs. \ref{fig:tp_spectrum}, \ref{fig:narrow_cmp_spectra} and \ref{fig:scale_par}. A direct quantitative confrontation of this model with our observed 0.1 
keV - 10 GeV phase-resolved spectra is therefore strongly recommended. \cite{grcrab_chengthree} compared in their paper the 
model calculations with phase-resolved Crab spectra from EGRET ($> 30$ MeV, Fierro et al. 1998). By mistake, EGRET spectra 
from the phase-resolved spatial analysis were used, which also comprise the underlying nebula component. This explains the 
large discrepancies between the model calculations and the observed spectra for energies between 30 and 100 MeV.
Cheng et al. (2000) also show a broad-band (0.1 keV - 10 GeV) model spectrum for the phase-averaged Crab pulsar spectrum, 
which can be compared with our spectrum in Fig. \ref{fig:tp_spectrum}. The overall shape of the model spectrum follows the 
observed spectral characteristics well, although for energies below $\sim 100$ keV the model underestimates the observed X-ray 
fluxes, i.e. it seems that the component which we empirically described as a narrow spectral ``bump'' peaking around 20 keV  
in an $E^2 \times F$ representation, is not accounted for. This can be best studied in the phase-resolved analysis.

Future observations of the Crab pulsar by high-energy missions are important. In particular the spectral characteristics in 
the 300-1000 keV interval must be determined much more accurately. Here, data from both IBIS and SPIE aboard INTEGRAL will 
contribute significantly. At high $\gamma$-ray energies data from the AGILE and GLAST missions can provide sufficient 
statistical precision to study the peculiar spectral behaviour in the LW2 phase interval in much more detail for energies above 1 GeV. 
Moreover, these instruments can, for the first time, study the pulsed emission for energies above 10 GeV (upper bound of the 
sensitivity window is 50 GeV and 300 GeV for AGILE and GLAST, respectively). 
At medium $\gamma$-ray energies there are plans for a mission with an advanced, more sensitive Compton telescope (called MEGA
0.5-50 MeV; see \cite{grcrab_kanbach}), but no approval for such a mission exists at this time. It is just in this energy range, 
where interesting spectral transitions occur and where we have indications for enhanced pulsed emission in the 10-15 MeV range. 
Future space borne Compton telescopes having 10-100 times better sensitivity than CGRO COMPTEL are required to allow further 
progress.


\begin{acknowledgements}
The COMPTEL project is supported by the German Ministerium f\"ur Bildung und Forschung 
through DLR grant 50 QV 90968 and by the Netherlands Organisation for Scientific Research 
(NWO). We are grateful to Jesper Sollerman, who kindly provided the HST STIS
pulse profiles of the Crab pulsar, and to Rudolf Much for making the optical UCL 
MIC Crab data accessible.
\end{acknowledgements}


\appendix

\section{Uncertainties in the absolute flux measurements at hard X-rays/soft $\gamma$-rays}
\label{app_flxcal}

In this work we compiled Crab spectra from soft X-rays up to high-energy gamma rays in order to better determine its spectral 
characteristics. Meanwhile, the Crab is used by many instruments as an in-flight calibration source. However, how well do we know 
the genuine Crab spectrum? At lower X-ray energies the measurement of the spectrum is coupled to an estimate of the 
N$_{\hbox{\rm \scriptsize H}}$ value, at harder X-rays up to the gamma-ray regime we have direct measurements 
of the Crab spectrum, but how realistic are the estimates of the accuracy of the pre-launch instrument calibrations? Is there any for
which we can trust the claimed accuracies most? In this Appendix we show the present status of our knowledge of the Crab spectrum. 
The still large (see below) systematic differences between different instruments should be kept in mind, not only when drawing 
conclusions on Crab results, but for any other source in high-energy astrophysics.

In Sect. \ref{section_opspec} we derived a N$_{\hbox{\rm \scriptsize H}}$ value of $3.61(2) \times 10^{21}\ \hbox{\rm cm}^{-2}$.
This estimate is significantly larger than the N$_{\hbox{\rm \scriptsize H}}$ value of $3.23(2) \times 10^{21}\ \hbox{\rm cm}^{-2}$ 
derived recently by Massaro et al. (2000) analysing a much larger BeppoSAX Crab database. 
The apparent discrepancy can be explained by their use of an older version of the LECS response description in combination with fitting LECS 0.1-4 keV data only, thus constraining to a lesser extent the photon index. We verified this by reproducing our values 
for N$_{\hbox{\rm \scriptsize H}}$, the normalization and the photon index for this Off Pulse emission repeating our analysis 
for the database used by Massaro et al. (2000).
From this comparison it is clear that the uncertainty in N$_{\hbox{\rm \scriptsize H}}$ is dominated by systematic uncertainties
in the response characteristics, of the LECS in particular, rather than by statistical ones.
 
Our value for N$_{\hbox{\rm \scriptsize H}}$ of $3.61(2)\times 10^{21}\ \hbox{\rm cm}^{-2}$ is consistent with the X-ray based 
value of $3.45(42) \times 10^{21}\ \hbox{\rm cm}^{-2}$ obtained by \cite{grcrab_schattenburg} using the Focal Plane Crystal Spectrometer on the Einstein observatory and with a radio based estimate of $\sim 3.65 \times 10^{21}\ \hbox{\rm cm}^{-2}$ 
(\cite{grcrab_dickey}). Recently, a study using XMM Newton EPIC MOS data yields a hydrogen column density of $3.45(2) \times 10^{21}\ 
\hbox{\rm cm}^{-2}$ for an oxygen-iron depleted abundance of $0.63(1)$ solar (\cite{grcrab_willingale}). Assuming solar abundances the 
N$_{\hbox{\rm \scriptsize H}}$ value lowers to $3.28(2)\times 10^{21}\ \hbox{\rm cm}^{-2}$. These estimates indicate that the genuine value 
of N$_{\hbox{\rm \scriptsize H}}$ will very probably lie in the range $(3.3-3.6) \times 10^{21}\ \hbox{\rm cm}^{-2}$.

{\begin{figure}[t]
  \hspace{0.3cm}
  {\psfig{file=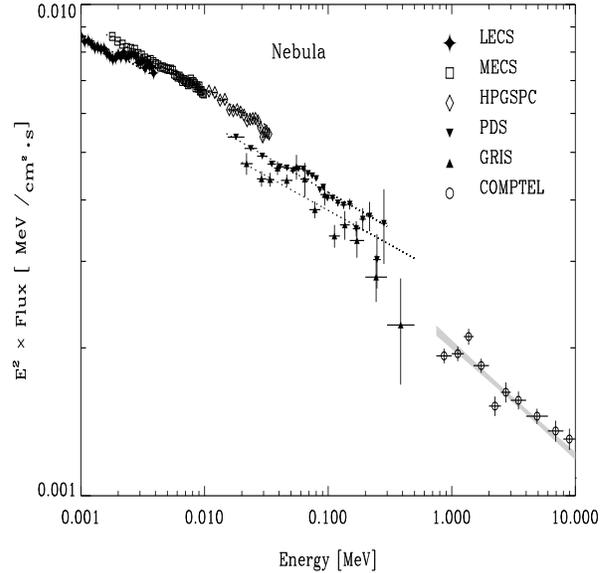,width=8cm,height=8cm}}
  \caption[]{The Crab nebula spectrum in the 1 keV - 10 MeV energy interval.
             Flux measurements from BeppoSAX LECS (1-4 keV), MECS (1.6-10 keV), 
             HPGSPC (10-32 keV) and PDS (15-300 keV), GRIS (20-500 keV) and COMPTEL 
             (0.75-10 MeV) are shown. The dotted lines show the best 
             power-law fits for the combined BeppoSAX 
             instruments and for GRIS. Note the apparent systematic deviation from the power-law fit 
             in the LECS spectral data. The shaded band indicates the $\pm 1\sigma$ uncertainty interval around 
             the optimum power-law fit for COMPTEL (0.75-30 MeV). 
             Clear discrete jumps are visible between the various BeppoSAX instruments and
             GRIS reflecting uncertainties in absolute sensitivity.
             \label{fig:op_zoomin}}
\end{figure}}

In the fit to the data of the 4 BeppoSAX NFI instruments we derived for the sensitivity normalization scale factors, relative to 
the MECS (factor set to 1): LECS 0.93, HPGSPC 1.01 and PDS 0.87. This means that the LECS and PDS calibrations of their overall 
sensitivities deviate from that of the MECS by $7\%$ and $13\%$, respectively.
This is clearly visible in Fig. \ref{fig:op_zoomin}, which presents the nebula spectrum in an $E^2\times F$ representation 
between 1 keV and 10 MeV as measured with the 4 BeppoSAX NFI instruments and COMPTEL, in which no normalization correction 
factors have been applied. Also shown is the Crab nebula spectrum in the 0.02-1 MeV energy range as measured by the balloon borne 
GRIS (Ge detectors; Bartlett et al. 1994a). The combined BeppoSAX spectra as well as the GRIS spectrum are best fitted with 
a power-law spectral shape with a consistent slope of $\sim 2.14$ over the 1-700 keV interval. However,
the normalization factor for GRIS, relative to the MECS is even as low as 0.78, to be compared with the
estimated GRIS systematic uncertainties of $+12\%$ and $-6\%$ by  \cite{grcrab_bartlettb}. Allowing spectral curvature in the 
multi-instrument BeppoSAX nebula fit by introducing an energy dependent power-law index does not improve the fit significantly. 
The same is true for the GRIS spectrum. However, some gradual softening above a few 100 keV is required
to connect to the softer spectrum measured by COMPTEL at energies above 1 MeV (cf. Fig. \ref{fig:op_zoomin}; power-law photon 
index in the 0.75-30 MeV interval is $2.227\pm 0.013$). 
The statistical uncertainties in the above quoted normalization correction factors are typically better than $1\%$. Therefore, 
it is obvious that the differences in absolute normalizations are systematic, and it is discouraging to note that between the two 
instruments for which the calibrations were expected to be most accurate (MECS and GRIS), the discrepancy appears to be largest. 
We cannot decide unambigously on this controversy. Therefore, we feel that in the presentation and analysis of the above spectral 
data in combined broad-band spectra, this problem should not be hidden by making arbitrary choices on normalization. 



\end{document}